\PassOptionsToPackage{sort&compress,numbers}{natbib}

 \documentclass[5p,times,twocolumn]{elsarticle}

\usepackage{amssymb}

\usepackage{float}
\usepackage{amsmath, amsfonts, amssymb, amsthm}
\usepackage{braket}
\usepackage{bbold}
\usepackage{dsfont}
\usepackage{graphicx}
\usepackage{url}
\graphicspath{ {./} }

\usepackage{wrapfig}
\usepackage{float}
\usepackage{pythontex} 
\usepackage{csvsimple}
\usepackage{pgfplotstable}
\usepackage{listings}
\usepackage{plain}
\usepackage{cleveref}
\usepackage{pdfpages}
\usepackage{makecell}

\usepackage[utf8]{inputenc}
\usepackage{amsmath}
\usepackage{graphicx}
\usepackage{subfig}
\usepackage{cancel}

\newcounter{bla}

\journal{Computer Physics Communications}

\begin{document}

\begin{frontmatter}





\title{jahn-teller-dynamics: Python package to determine vibronic interaction demonstrated on molecules and trigonal defect qubits}

\author[a,b]{Bal\'azs T\'oth}
\author[b,a,c]{\'Ad\'am Gali\corref{author1}}
\author[b]{Gerg\H{o} Thiering\corref{author2}}

\cortext[author1] {Corresponding author.\\\textit{E-mail address:} gali.adam@wigner.hun-ren.hu}
\cortext[author2] {Corresponding author.\\\textit{E-mail address:} thiering.gergo@wigner.hun-ren.hu}
\address[a]{Department of Atomic Physics, Institute of Physics, Budapest University of Technology and Economics, M\H{u}egyetem rakpart 3., H-1111 Budapest, Hungary}
\address[b]{HUN-REN Wigner Research Centre for Physics, Institute for Solid State Physics and Optics, P.O.\ Box 49, H-1525 Budapest, Hungary}
\address[c]{MTA-WFK Lend\"ulet "Momentum" Semiconductor Research Group, P.O.\ Box 49, H-1525 Budapest, Hungary}

\begin{abstract}
Trigonal solid-state defects are often subjects of spontaneous symmetry breaking driven by the $E\otimes e$ Jahn-Teller effect, reflecting strong electron-phonon coupling.
These systems, particularly paramagnetic defect qubits in solids are central for quantum technology applications, where accurate knowledge of their fine-structure parameters – shaped by the complex interplay of spin-orbit and electron-phonon interactions – is essential.
We introduce the $\mathtt{jahn {\text -} teller {\text -} dynamics}$ package, a Python code that implements the first-principles approach of [Phys. Rev. X 8, 021063 (2018)] to accurately compute the spin-orbit-phonon entanglement in trigonal defects utilizing the output from density functional theory calculations (DFT) to predict fine-structure parameters of zero-phonon lines (ZPLs), including Zeeman shifts under external magnetic fields.
We demonstrate its capabilities on negatively charged Group-IV–vacancy (G4V) defects in diamond: SiV$^-$, GeV$^-$, SnV$^-$, PbV$^-$ and the neutral N$_3$V$^0$ defect in diamond, and the CH$_3$O$^0$ methoxy radical.
Additionally, we implement a generic electron-phonon code that is capable entangling an arbitrary amount of (i) vibration modes and (ii) electronic levels by (iii) arbitrarily high order of vibronic interaction terms.
Exemplarily, we demonstrate the Jahn-Teller multimode problem on the CH$_3$O$^0$ methoxy radical and the Pseudo Jahn-Teller case on the $C_4H_4^+$ butatrien cation.
\end{abstract}


\end{frontmatter}



{\bf PROGRAM SUMMARY/NEW VERSION PROGRAM SUMMARY}

\noindent
{\em Program Title: jahn-teller-dynamics}                                          \\
{\em CPC Library link to program files:} (to be added by Technical Editor) \\
{\em Developer's repository link:}\\
https://github.com/tbalu98/Jahn-Teller-Dynamics \\
https://pypi.org/project/jahn-teller-dynamics \\
{\em Code Ocean capsule:} (to be added by Technical Editor)\\
{\em Licensing provisions(please choose one):} GPLv3  \\
{\em Programming language: Python 3.11}                                   \\
{\em Data availability: https://doi.org/21.15109/ARP/EXJKGL}                                 \\
{\em Nature of problem(approx. 50-250 words):}\\
Electron states that have close or degenerate or degenerate energy levels are susceptible to atomic vibrations caused by atomic nuclei that causes the dynamic Jahn-Teller and pseudo Jahn-Teller interaction. Indeed, density functional theory (DFT) and multiconfigurational Hartree-Foch calculations within the Born-Oppenheimer approximation can be employed to determine the electronic structure in different geometrical configurations or determine the spin-orbit coupling strength.
However, in order to solve Jahn-Teller problem one has to go beyond the Born-Oppenheimer approximation that is usually well beyond the capabilities of current state-of-art codes.
Therefore, a first-principles approach is necessary to accurately determine the fine-structure parameters such as the spin-orbit splitting visible in optical measurements utilizing trigonal color centers and molecules.
 
{\em Solution method(approx. 50-250 words):}\\
In $\mathtt{jahn-teller-dynamics}$ library we implemented a general formulation of dynamic Jahn-Teller theory, where arbitrary number of electron states and phonon modes interact with each other up to any order. Therefore, the user can freely customize the Hamilton operator of the electron-phonon interaction. We dedicated extra care to the $E\otimes e$ dynamic Jahn-Teller interaction, that is very common among point defects. We implemented methodology developed in Refs.~\cite{Thiering2017, PhysRevX2018}. It utilizes the results directly obtained from the $\mathtt{VASP}$\cite{vasp} density functional theory (DFT) code to formulate the Hamiltonian of the $E\otimes e$ Jahn-Teller case.
Our solution includes the non-perturbative effect of spin-orbit coupling and the dynamic Jahn-Teller interaction simultaneously, where we observe that these two interactions non-trivially entangle with each other.
Our methodology determines the damping of experimentally visible spin-orbit splitting known as Ham~\cite{Ham_red_fact, Bent_1990, Norambuena_2020} reduction factor $p$ that can be identified as electron-phonon renormalization of physical observables such as the renormalization of spin-orbit coupling strength~\cite{Csore2020, Thiering_2024, PhysRevB.110.184302, thiering2024nuclear} that of trigonal defects.
Our code provides a framework that automatically reads and post-processes the DFT results to compute fine-structure parameters and it 
includes the effect of external magnetic fields.


\section{Introduction}

Point defects, such as vacancies, dopants and defect complexes in semiconductors and insulators such as diamond and silicon carbide hold great promise to realize solid-state quantum bits~\cite{Weber_2010, Atat_re_2018, Gali2020ApplPhysRev, Bassett_2019, Chatterjee_2021, Wolfowicz_2021}. The optically active quantum bit candidates are fluorescent exhibiting coherent emission within their zero-phonon-line (ZPL) which play a crucial role in setting, manipulating and readout of defect-based quantum bits fully optically~\cite{Doherty2013, Gali2019, sil_vac, GeV_exp2015, GeV_exp2017, tin-vac_diamond, tin_fine_structure, Gorlitz_2020, PbV_exp}. The underlying defect quantum bits often exhibit high symmetry, giving rise to degenerate orbitals that couple to spin, resulting in fine-structure splitting, and interact strongly with phonons also known as vibronic states. According to dynamic Jahn-Teller theory, which goes beyond the standard adiabatic approximation~\cite{DJT_theory, LonguetHiggins1961, Bersuker_2006, bersuker2012vibronic}, the strong electron-phonon interaction can quench orbital angular momentum and thereby reduce the effective spin-orbit coupling~\cite{Thiering2017,PhysRevX2018}.

In this work, we introduce  $\mathtt{Exe.py}$, an open-source Python code designed to simulate the fine structure parameters governing ZPL transitions of these defects under applied magnetic fields, using input from density functional theory (DFT) calculations. We note that our scheme can be applied to dynamic Jahn-Teller active trigonal molecules mostly exhibiting the $X    \mathrm{CH_3}$ structure where $X=$~Cd, Zn, Mg, Ca, O, Sr, etc.  \cite{Momose_1988, Momose_1989, Bent_1990, Brazier_1989, Barckholtz_1999,  Melnik_2011, Liu_2009, Martin_Drumel_2025, Sharma_2025, Nagesh_2014, Johnson_2017, Sharma_2019, Knight_1991, Dick_2006}. The code is developed using widely adopted scientific libraries such as NumPy, Pandas, SciPy, and Matplotlib. It implements the formalism presented in Refs.~\cite{Thiering2017, PhysRevX2018}, allowing direct comparison with experimental observations.

\begin{figure}[h]
\includegraphics[width=0.45\textwidth]{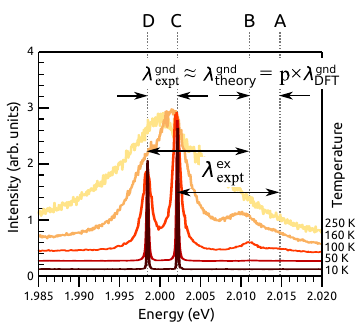}
\centering
\caption{Relationship between the experimentally observed photoluminescence spectrum of $\text{SnV}^-$  and the theoretical value of spin-orbit coupling $\lambda^{\text{gnd, ex}}_{\text{expt, theory}}$ of the ground (gnd) and excited (ex) states. The data for the experimental spectrum is taken from Ref.~\cite{Gorlitz_2020}.}
\label{fig:exp_th}
\end{figure}
Notable examples of such defects include nitrogen-vacancy center~\cite{Doherty2013, Gali2019} and group-IV--vacancy centers~\cite{PhysRevX2018} in diamond or silicon-vacancy~\cite{Janzen2009, Baranov2011, Dyakonov2013, Widmann2015} and divacancy centers~\cite{SonPRL2006, Gali2011, KoehlNature2011, SiC_divac} in silicon carbide where other host materials with defects~\cite{Gali2020ApplPhysRev} and molecular systems~\cite{AwschalomScience} have recently emerged as quantum bit candidates.
Exemplarily, we compare the experimental ZPL spectrum that of $\text{SnV}^-$ \cite{Gorlitz_2020} with our theoretically predicted spin-orbit coupling fine structure parameters ($\lambda^{\text{gnd/ex}}_{\text{expt/theory}}$). Fig.~\ref{fig:exp_th}. represents how they can be measured from the photoluminescence spectrum. The loss of degeneracy due to spin-orbit coupling manifests in ZPL splitting. Indeed there are four possible transition energies between the ground and excited state, however in low temperatures only the lowest orbital states are populated, therefore we anticipate two peaks. Higher orbital states are populated at higher temperatures (100K), but they are hard to resolve because of the increasing intensity of the phonon bath.

\section{Basic theory}
\label{sec:theory}
Most generally, one may consider the following $\hat{H}$ Hamiltonian 
\begin{equation}
    \label{eq:generic_e_ph1}
    \hat{H}=\hat{H}_{\text{e}}+\hat{H}_{\text{ph}}+\hat{H}_{\text{e-ph}} \text{,}
\end{equation}
that depicts the interaction between electronic ($\hat{H}_{\text{e}}$) and vibrational ($\hat{H}_{\text{vib}}$) degrees of freedom spanned by a set of many-electron wavefunctions ($|n\rangle$) indexed by $n$ labels exhibiting $\varepsilon_{n}$ eigenenergies and a set of quantum harmonic oscillators indexed by $i$ labels depicted by their creation $a_i^\dagger$ and annihilation $a_i$ ladder operators exhibiting $\omega_{i}$ energy quanta, respectively:
\begin{equation}
    \label{eq:generic_e_ph2}    \hat{H}_{\text{e}}=\sum_{n}\varepsilon_{n}|n\rangle\langle n|\text{,}\qquad\hat{H}_{\text{vib}}=\sum_{i}\omega_{i}\left(a_{i}^{\dagger}a_{i}+\frac{1}{2}\right)\text{.}
\end{equation}

The occupation number representation of the N-mode quantum harmonic oscillator can be introduced by consecutive creation of phonon quanta on the vacuum state:
\begin{equation}
    \begin{split}
        |\nu_{1},\nu_{2},..,\nu_{N}\rangle=\prod_{i}\biggl(\frac{1}{\sqrt{\nu_{i}!}}(a_{i}^{\dagger})^{\nu_{i}}\biggr)|0,0,..,0\rangle\text{.}
    \end{split}
\end{equation}

The resulting basis states of the Hilbert space $\mathcal{H}$ are 

\begin{equation}
    b_j = \ket{n, \nu_1, \nu_2, ..., \nu_N}
\end{equation}

The interaction Hamiltonian ($\hat{H}_{\text{e-vib}}=\sum_{nm}|n\rangle\langle m|\hat{H}^{nm}_{\text{e-vib}}$) is a generic series expansion made of $\hat{q}_{i} =\frac{1}{\sqrt{2}} (a_i^\dagger+a_i)$ coordinate operators and $|n\rangle\langle m|$ many-electron excitation operators\cite{Sharma_2025}:
\begin{equation}
    \label{eq:generic_e_ph3}
    \begin{split}
  \hat{H}^{nm}_{\text{e-vib}}
  &=\sum_{\alpha}F_{\alpha}^{nm}\hat{q}_{\alpha}
+\sum_{\alpha\beta}F_{\alpha\beta}^{nm}\hat{q}_{\alpha}\hat{q}_{\beta}\\
&+\sum_{\alpha\beta\gamma}F_{\alpha\beta\gamma}^{nm}\hat{q}_{\alpha}\hat{q}_{\beta}\hat{q}_{\gamma}
+\sum_{\alpha\beta\gamma\delta}F_{\alpha\beta\gamma\delta}^{nm}\hat{q}_{\alpha}\hat{q}_{\beta}\hat{q}_{\gamma}\hat{q}_{\delta} + ...
    \end{split}
\end{equation}
The matrix elements describe the coupling between the $|n\rangle$ and $|m\rangle$ electron states as a Taylor series expansion up to an arbitrary order in terms of the normal modes $\hat{q}_\alpha$. $F^{nm}_{\alpha...} = \frac{1}{k!}f^{nm}_{\alpha...}$, where $f^{nm}_{\alpha...}$ is the kth order derivative of the adiabatic potential energy surface $V(\hat{q}_1, \hat{q}_2, ...)$. $F^{nm}_{\alpha\beta\ldots}$ for all permutations of indexes $\alpha\beta...$ coefficients are equal.

The position operators can be dimensionless $\hat{x}_i = \frac{1}{\sqrt{2}} (\hat{a}^\dagger + \hat{a})$ or dimensional $\hat{x}_i = \frac{1}{\sqrt{2\omega_i}} (\hat{a}^\dagger + \hat{a})$. $\omega_i$ is the energy quantum of the ith vibrational mode, $\hat{a}^\dagger_i$ is the creator and $\hat{a}_i$ is the annihilator operator.

This general Hamiltonian operator can be built by \texttt{jtd\_run} program. Example files of user inputs are provided in the github repository along with a technical description. Here we aim to cover the Hamiltonian building principles of our code. The code requires the user to describe $N_e$ electron states $\varepsilon_{n}$  energy levels and the vibration energy quantum of the vibrational modes $e_i$. \texttt{jtd\_run} code can build tensor product phonon states with respect to a maximum phonon quanta number on each modes $N_i$ such that $\ket{\nu_1, \nu_2, \ldots } $ such that for each vibrational quanta number $\nu_i \leq N_i$ . The other option is that the user can define the maximum energy $(E)$ of the phonon basis, such that $\sum_i  \nu_i \omega_i \leq E$,or the maximum number of phonon quanta $N_{\text{ph}}$, such that $\sum_i \nu_i \leq N_{\text{ph}}$

\subsection{Hamiltonian building principles of \texttt{jtd\_run}}

Due to symmetry considerations for any arbitrary order k not all monomials have distinct coupling coefficients. For instance in case of $E\otimes e $ dynamic Jahn-Teller interaction, all second order monomials couples to the electron states with the same coefficients. Therefore, the coupling between two electron states $\ket{n}$ and $\ket{m}$ can be defined by arbitrary $p_{nm,k_l}$ polynomial consists of $k$th order monomials with a coupling coefficient $c_{p^{nm}_{k_l}}$. The kth order polynomials are indexed by $l$ in order to take into account all the kth order polynomial that couples $\ket{n}$ and $\ket{m}$. \texttt{jtd\_run} can be configured to create a permutation completion, which means that it automatically generate all distinct permutation of $p_{nm,k_l}$'s monomials. In this case of a tensor product phonon basis, position operators commute. Therefore, the monomials can simply be multiplied by the number of unique permutations. Otherwise all distinct monomials are generated, because the position operators do not commute. The resulting new polynomial noted by $p'_{nm,k_l}$.

To concretely exemplify it, in case of a coupling polynomial $x^3 + x^2y$ for non commuting and commuting cases:

\begin{equation}
        p'^{nm}_{ k_l} = x^3 + x^2y + yxy + y^2x = x^3 + 3x^2y
\end{equation}

The code iterates over the user defined input and generates the Hamiltonian operator based on this sum:

\begin{equation}
\label{eq:H_code_sum}
    \begin{split}
        \hat{H}_{\text{built}} = \sum_{nm,k_l} c^{nm}_{k_l} \ket{n}\bra{m}\otimes p'^{nm}_{k_l}
    \end{split}
\end{equation}

Here $\ket{n}\bra{m}$ is an $N_e\times N_e$ matrix. $\ket{n}\bra{m}_{\kappa \lambda} = 1.0$, if $\kappa=n$ and $\lambda=m$, $0.0$ otherwise.

For simplicity, we encourage users to define only the upper triangle of the Hamiltonian because Hermitian completion option can be set. Therefore the code automatically performs the summation \eqref{eq:H_code_sum}, and includes the hermitian conjugate of each term.

In some cases it can take long time to generate all coupling polynomials, the user can choose to opt out Hermitian completion and set symmetrization such that at the end of the Hamitlonian building procedure $\hat{H}_{\text{e-ph}} = \frac{1}{2} \left(\hat{H}_\text{built} + \hat{H}_\text{built}^\dagger\right)$. Note that in that case the non diagonal coupling coefficients must be doubled.

To sum up the component of the user defined coupling coefficients:

\begin{equation}
    \begin{split}
        c^{nm}_{k_l} = \underbrace{2\times \underbrace{ \dfrac{1}{k!}\times f^{nm}_{k_l}}_{\text{Herm. comp.}} }_{\text{symmetrize Hamiltonian}}
    \end{split}
\end{equation}

Here $f^{nm}_{k_l}$ is the adiabatic potential energy surface derivative with respect to the coordinates of any monomial in $p_{k_l}$

\begin{figure}[H]
\includegraphics[width=\columnwidth]{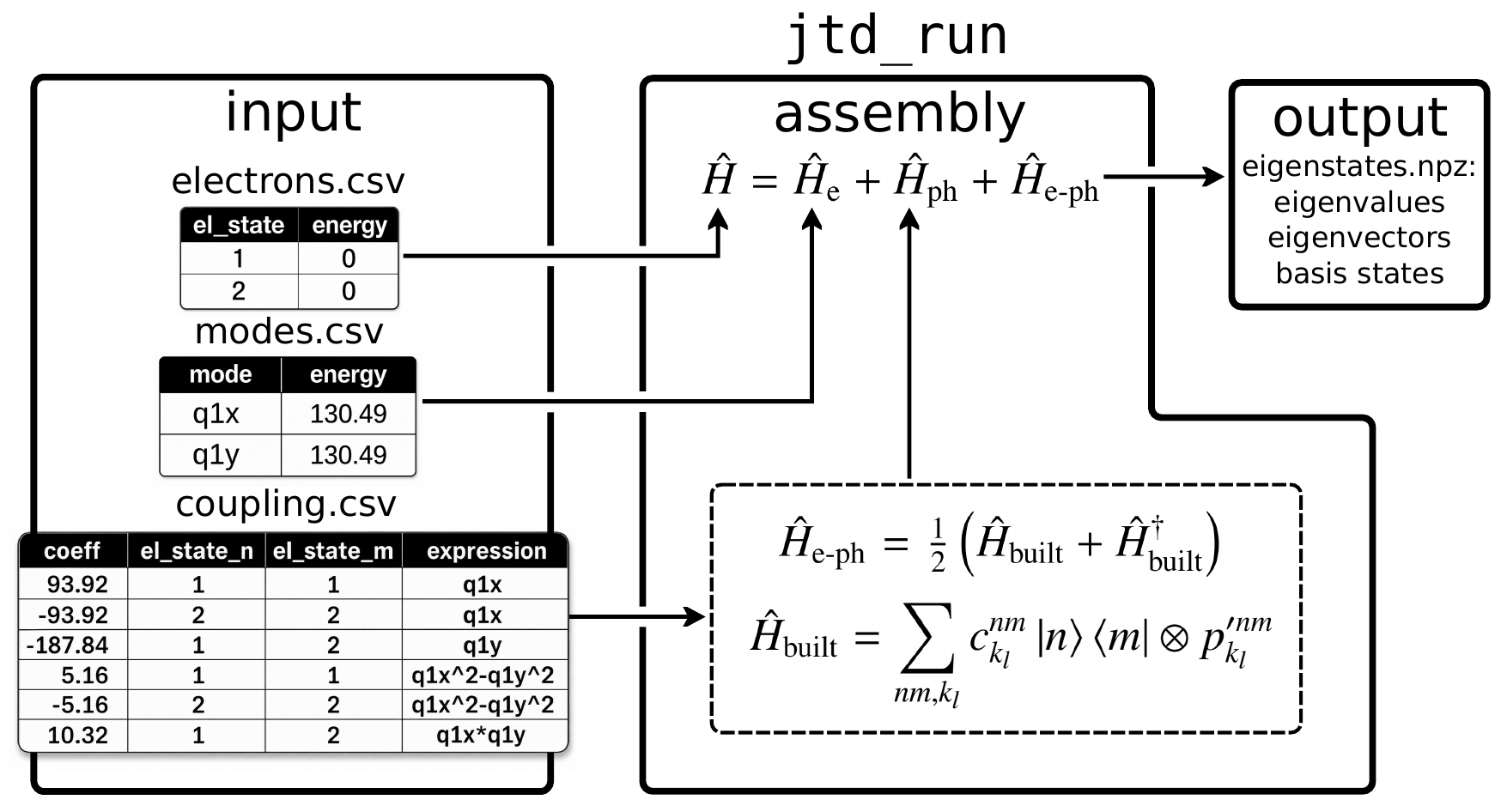}
\caption{\texttt{jtd\_run} reads electron energies, phonon modes and coupling defined in a tabular form. It builds the general Hamiltonian \ref{eq:gen_H} and returns the eigenvalues and eigenvectors, that are meant in terms of the basis states}
\centering
\label{fig:jtd_io}
\end{figure}

Moreover, the code can optionally diagonally complete the Hamiltonian. Therefore, when the user specifies one diagonal element for an electron state, $i$ it generates that input for all other states as well. Figure \ref{fig:jtd_io} presents the operating principles of \texttt{jtd\_run}.

\texttt{jtd\_run} uses Lanczos algorithm as implemented in \texttt{scipy} and \texttt{SPLEPc} in order to find the first $N_{\text{eigen}}$ eigenvalues and eigenvectors in terms of the basis states required by the user and saves in \texttt{.npz} format, named \texttt{eigenstates.npz}. It contains the basis states of Hilbert space $(\mathcal{H})$ in (\texttt{basis\_labels.npy}), the eigenvalues of the dynamic Jahn-Teller system $E_\eta$ (\texttt{eigenvalues.npy}) and the eigenstates $\psi_\eta$(\texttt{eigenvectors.npy})

\subsection{Single mode $E\otimes e$ Jahn-Teller case for defects in solids}
The general purpose of $\mathtt{jahn {\text -} teller {\text -} dynamics}$ is to evaluate the following interaction Hamiltonian ~\cite{ PhysRevX2018, Bersuker_2006, Wang_2019}
\begin{equation}
    \label{eq:gen_H}
    \hat{H} = \hat{H}_{\text{SOC}}+ \hat{H}_{\text{vib}} + \hat{H}_{\text{DJT}}  + \hat{H}_{\text{ext}} \text{,}
\end{equation}
to theoretically predict the fine structure details for defects such as the $\lambda_\mathrm{theory} \approx \lambda_\mathrm{expt}$ spin-orbit splitting parameter visible in experiments. 
In the following sections, we describe the terms of Eq.~\eqref{eq:gen_H}. First, we discuss the intrinsic spin-orbit coupling $\hat{H}_{\text{SOC}}$  in Sec.~\ref{sec:SOC}, that we determine by means of DFT calculations. 
Next, we introduce the dynamic Jahn-Teller effect in Sec.~\ref{sec:DJT} where  $\hat{H}_{\text{vib}}$ depicts the atomic vibrations that are represented by quantum harmonic oscillators while $\hat{H}_{\text{DJT}}$ is the electron-phonon interaction term due to the dynamical Jahn-Teller effect. Finally, the code directly diagonalizes $\hat{H}$ yielding the theoretical spin-orbit-phonon spectrum (see Sec.~\ref{sec:SOCJT} for details). The last term ($\hat{H}_{\text{ext}}=\hat{H}_{\text{mag.}}+\hat{H}_{\text{strain}}$) depicts the coupling of electronic orbital and spin with external magnetic field, electric field and strain that we discuss in Section~\ref{sec:Bfield}. 

\begin{figure}[h]
\includegraphics[width=0.45\textwidth]{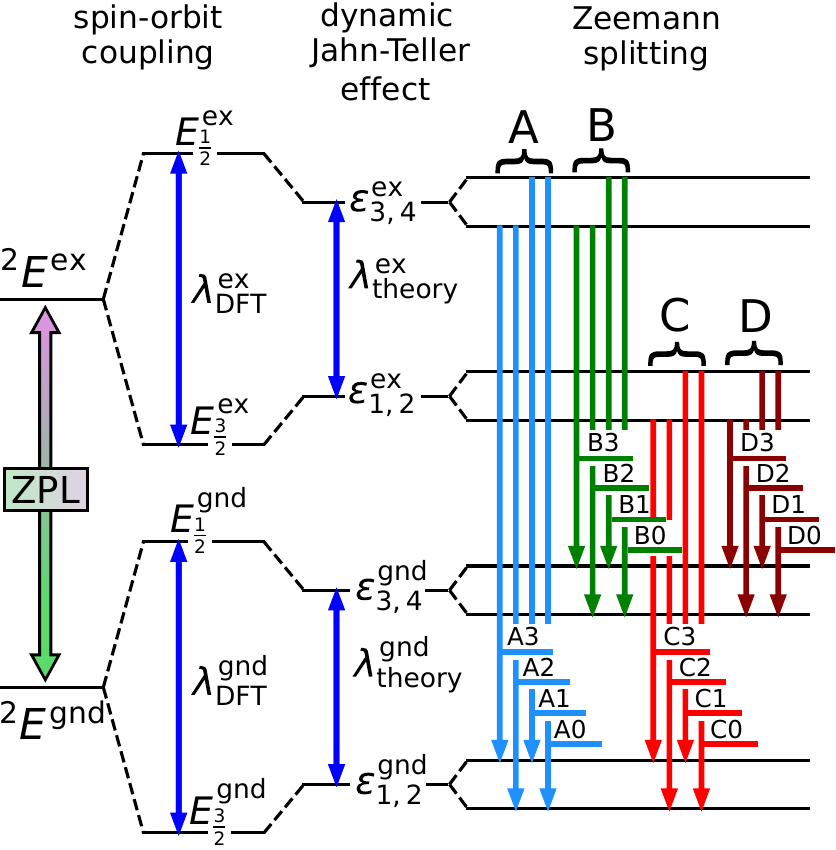}
\centering
\caption{This figure represents the physics behind the zero-phonon-line (ZPL) transitions in trigonal point defects with degenerate electronic ground and excited states. The spin-orbit electronic states of the ground and excited state are fourfold degenerate. Due to spin-orbit coupling they split into two $E_{\frac{1}{2}}$ and $E_{\frac{3}{2}}$ multiplets and the energy difference between them is $\lambda_0$. If we take into consideration atomic vibrations by utilizing the $E\otimes e $ Jahn-Teller effect case, they will split to two distinct two-fold degenerate energy levels $(\ket{\varepsilon_{1}}, \ket{\varepsilon_{2}})$ and $(\ket{\varepsilon_{3}}, \ket{\varepsilon_{4}})$. They are entangled states of phonons, orbitals and spin. Under a constant external magnetic field the system experiences total loss of degeneracy because of Zeeman-effect. A,B,C,D represents the transition types as defined in Ref.~\cite{tin_fine_structure}.}
\label{fig:sketch}
\end{figure}

We note that the fine structure parameters can be encompassed into an effective model Hamiltonian for the vibronic ground state where the vibrational degrees of freedom ($X$, $Y$) can be traced out~\cite{PhysRevX2018, Norambuena_2020} by averaging over the phonon bath.
We discuss the properties of the effective Hamiltonian in Sec.~\ref{sec:modelHamilton} where the intrinsic $\lambda_0$ spin-orbit parameter is renormalized into an effective $\lambda_{\text{theory}}=p\times \lambda_0\approx \lambda_{\text{expt}} $ energy gap visible in experiments through the introduction of Ham reduction factors.
Whenever both ground and optical excited states are dynamic Jahn-Teller active multiplets, then Eq.~\eqref{eq:gen_H} can be applied for both states to determine the fine structure features in the optical spectra.
In such systems, $\mathtt{jahn {\text -} teller {\text -} dynamics}$ can simulate the fine levels in the optical spectrum as schematically depicted in Figs.~\ref{fig:sketch} and \ref{fig:tin_ZPL}.

Finally, we exemplarily depict the workflow with our code on experimentally known defects such as G4V defects (SiV, GeV, SnV, PbV) in Sec.~\ref{sec:numericalDFT}  including computational details (Sec.~\ref{sec:compdet}), where we highlight the $\text{N}_3\text{V}^0$ defect (Sec.~\ref{sec:N3V}) whose spin-orbit parameter has not been determined by ab-initio tools and compared to experimental data to our best knowledge.

\subsection{Trigonal system and spin-orbit coupling}
\label{sec:SOC}
We investigate point defects in solids exhibiting trigonal symmetry and electronically degenerate states. 
According to Jahn-Teller theory, such degenerate many-body electronic states undergo spontaneous symmetry breaking due to geometrical distortions.
In this context, adiabatic approximation  generally breaks down for degenerate (or quasi-degenerate) many-body electronic defect states in solids, owing to strong electron-phonon interaction.
In this work, we focus on defects exhibiting trigonal point-group symmetries such as $D_{3d}$ and $C_{3v}$ that can feature double degenerate orbitals or phonons labeled as $"E"$ or $"e"$ in Sch\"onflies notation. 
For the $D_{3d}$ point group, parity labels are omitted, as they hold no relevance in the present context.

The doubly degenerate orbitals $(\ket{e^{\text{DFT}}_{x}},\ket{ e^{\text{DFT}}_{y}})$ of point defects in solids or molecules can be approximately well characterized by DFT methodology.
When these degenerate Kohn-Sham orbitals are occupied by a single electron or equivalently left with a single hole results in a doubly degenerate many-body state. 
Furthermore, we note that taking into account the Kramers degenerancy of spin-$\frac{1}{2}$ systems yields a combined spin-orbital system exhibiting fourfold degeneracy.
By performing non-collinear calculations with spin-orbit effects turned on~\cite{PhysRevB.93.224425}, one can determine the spin-orbit splitting of these degenerate Kohn-Sham orbitals by setting half-half occupation of the degenerate Kohn-Sham levels~\cite{Thiering2017, PhysRevX2018}.
As a consequence, the Kohn-Sham states will acquire complex-valued character:
\begin{equation}
    \label{eq:complex_basis}
    \ket{e^{\text{DFT}}_{\pm}} =\mp\big( \ket{e^{\text{DFT}}_x}\pm i\ket{e^{\text{DFT}}_y}\big) \text{,}
\end{equation}
where the phases are chosen in accordance with the Condon-Shortley convention.
Their energy levels are separated by
\begin{equation}
    \lambda_0 = E\left(e^{\text{DFT}}_{+}\right)- E\left(e^{\text{DFT}}_{-}\right) \text{,}
\end{equation}
where $E\left(e^{\text{DFT}}_{\pm}\right)$ denote the eigenenergies of the corresponding orbital states.
One may observe that the $\hat{L}_{z} =  i  \left(\begin{smallmatrix}
		0 & -1\\
		1 & 0
	\end{smallmatrix}\right)$ assumes a simpler diagonal form upon transformation to the complex basis of Eq.~\eqref{eq:complex_basis}: $\hat{L}_{z, \text{complex}} =  \left(\begin{smallmatrix}
		1 & 0\\
		0 & -1
	\end{smallmatrix}\right)$ which explicitly reflects $\ket{e^{\text{DFT}}_{\pm}}$'s the orbital angular momentum quantum number $m_l = \pm 1$.
The spin states are $\ket{\uparrow}$ and $\ket{\downarrow}$, and the spin operator is $\hat{S}_z = \frac{1}{2} \left( \begin{smallmatrix}1 & 0\\0 & -1\end{smallmatrix}\right)$. The spin-orbit interaction operator is
\begin{equation}
	\hat{H}_{\text{SOC}} =  \lambda_0\hat{ L}_z\otimes \hat{S}_z \text{.}
\label{Eq.TheSOC}
\end{equation}
As a result, we find that the fourfold degenerate electronic states split into two Kramers doublets:
\begin{equation}
\begin{split}
    \ket{E_{\frac{1}{2}}} &= \{\ket{e^{\text{DFT}}_{+}}\otimes\ket{\downarrow}\text{, }  \ket{e^{\text{DFT}}_{-}}\otimes\ket{\uparrow}\} \text{,}\\
    \ket{E_{\frac{3}{2}}} &= \{\ket{e^{\text{DFT}}_{-}}\otimes\ket{\downarrow} \text{, }  \ket{e^{\text{DFT}}_{+}}\otimes\ket{\uparrow}\} \text{.}
\label{eq:DFT_E12_E32}
\end{split}
\end{equation}
where the spin–orbit interaction shifts $\ket{E_{\frac{1}{2}}}$ upwards while $\ket{E_{\frac{3}{2}}}$ downwards in energy by $\lambda_0/2$. 
However, the computed $\lambda_0$ substantially overestimates the experimentally measured $\lambda_{\text{expt}}$ for most defect-based quantum bits~\cite{Thiering2017, PhysRevX2018}. 
This discrepancy arises from strong electron-phonon coupling: where $\lambda_0$ represents only the spin-orbit splitting of the unrenormalized (bare) electronic states and is subsequently reduced by the Jahn-Teller effect, as discussed in the following section.

\subsection{$E \otimes e$ Jahn-Teller effect}
\label{sec:DJT}

In this section, we briefly outline how strong electron-phonon coupling in trigonal systems can partially quench the spin-orbit interaction and thereby substantially modify the effective spin-orbit splitting.
Within geometry optimization workflows during DFT calculations, one may observe that low symmetry configurations (e.g. $C_{2h}$ or $C_{1h}$) are energetically favored over the high-symmetry configuration for trigonal defects. In such distorted cases $\ket{e^{\text{DFT}}_x}$ and $\ket{e^{\text{DFT}}_{y}}$ orbitals become non-degenerate as the phonons perturb electronic structure.
These trigonal systems can be described within the framework of the $E \otimes e $ Jahn-Teller case as explained in Ref.~\cite{Bersuker_2006}. 
\begin{figure}[H]
\includegraphics[width=\columnwidth]{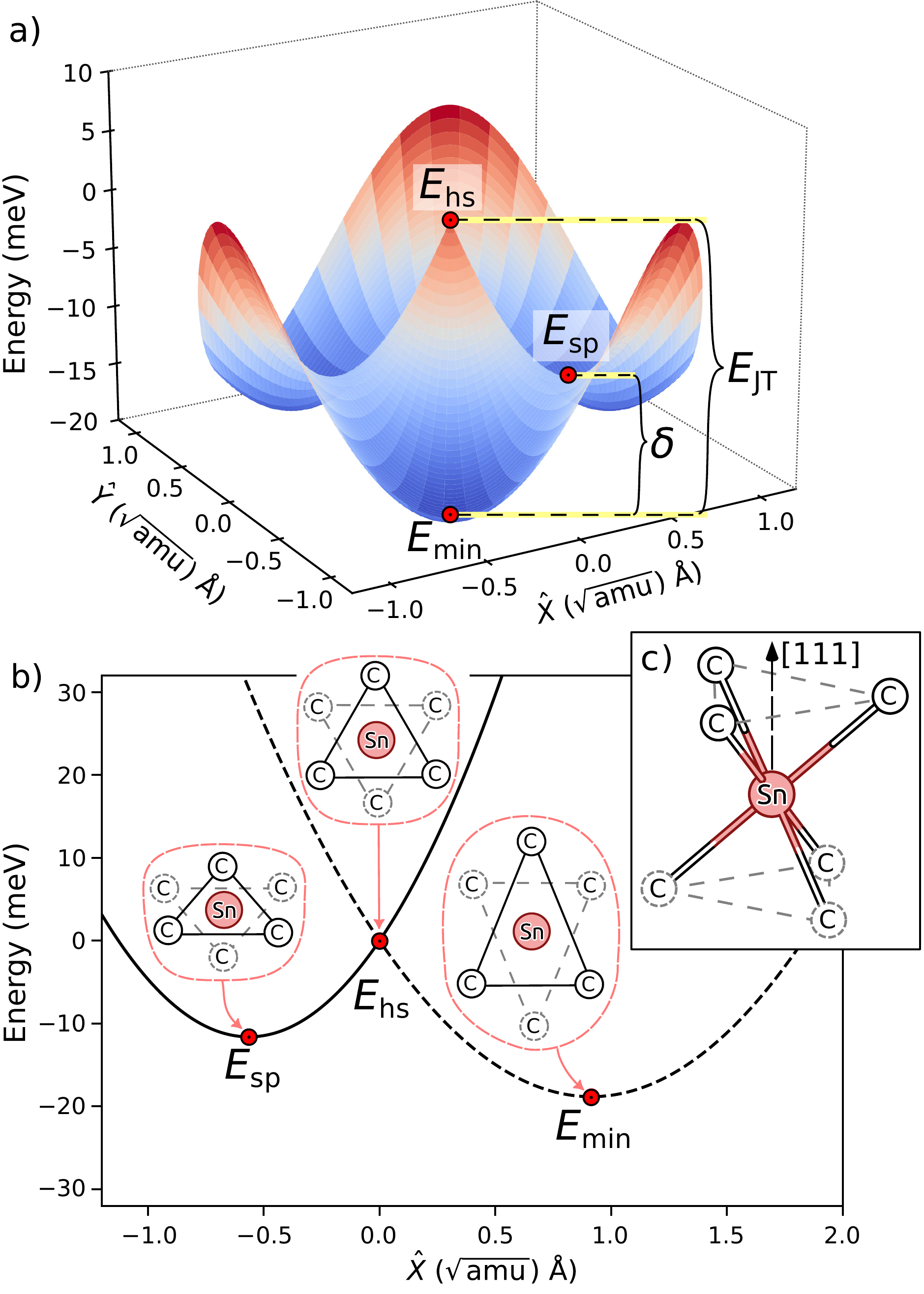}
\caption{(a) Adiabatic potential energy surface (APES) of the Jahn-Teller distortion. A one-dimensional cut through the APES surface is shown in (b). Panel (c) displays the atomic configuration of the SnV center in diamond, with the distortions indicated in the insets that of panel (b). By applying the same distortion in two opposite directions, the global energy minimum and saddle point configurations can be found.}
\centering
\label{fig:APES}
\end{figure}
The localized vibration ($E$) is also doubly degenerate, similarly to the orbital manifold, and thus it can be introduced as a two-dimensional harmonic oscillator: 
\begin{equation}
    \hat{H}_{\text{vib}} = \omega\big( a_X^\dagger a_X + a_Y^\dagger a_Y + 1 \big) \text{,}
\label{eq:vib}
\end{equation}
where $a_{X,Y}^\dagger$ and $a_{X,Y}$ are phonon creation and annihilation operators while $\omega$ is the energy quantum of the oscillator. In an electron-phonon entangled manifold, the vibrational Hamiltonian is required to be extended to $\hat{H}_{\text{vib}} \rightarrow \hat{H}_{\text{vib}}\otimes \mathds{1}_{\text{orb}}$
where the additional identity operator $\mathds{1}_{\text{orb}} = \left( \begin{smallmatrix}
    1&0 \\ 0&1
\end{smallmatrix} \right)$ spans the orbital degrees of freedom to the coupled system. However, from now on, we will neglect the trivial $\otimes$ and $\mathds{1}_\text{orb}$ for simplicity.
The two dimensional harmonic oscillation of the lattice couples to the doubly degenerate electron orbital states which can be described by the following Hamiltonian:
\begin{equation}
\label{eq:H_DJT}
\hat{H}_{\text{DJT}}=F(\hat{X}\hat{\sigma}_{z}-\hat{Y}\hat{\sigma}_{x})+G\bigl[(\hat{X}^{2}-\hat{Y}^{2})\hat{\sigma}_{z}+2\hat{X}\hat{Y}\hat{\sigma}_{x}\bigr]\text{,}
\end{equation}
where $\hat{X}$ and $\hat{Y}$ are the position operators of the vibrational mode with energy quantum $\omega$ while $\hat{\sigma}_z = |e_{x}\rangle\langle e_{x}|-|e_{y}\rangle\langle e_{y}|=\left( \begin{smallmatrix}
    1 & 0 \\ 0 & -1
\end{smallmatrix} \right)$ and $\hat{\sigma}_x =|e_{x}\rangle\langle e_{y}|+|e_{y}\rangle\langle e_{x}| =\left( \begin{smallmatrix}
    0 & 1 \\ 1 & 0
\end{smallmatrix} \right)$ Pauli matrices govern the orbital degrees of freedom. The quantum states of the system thus consist of doubly degenerate orbitals entangled with two-dimensional vibrational motion. We numerically diagonalize the $\hat{H}_{\text{vib}}+\hat{H}_{\text{DJT}}$ Hamiltonian.
The eigenstates of the coupled system can be expanded in the following series expansion: 
\begin{equation}
\label{eq:waveansac}
|\kappa_{i}\rangle=\sum_{k,l}\bigl(c_{k,l}^{(i)}|e_{x}\rangle+d_{k,l}^{(i)}|e_{y}\rangle\bigr)\otimes|k,l\rangle \text{,}
\end{equation}
where $\ket{e_{x,y}}$ are electronic orbitals while $\ket{k,l}$ are vibrational wavefunctions.
The latter represent the Fock states of the two-dimensional quantum harmonic oscillator along $X,Y$ directions, explicitly given by $|k,l\rangle=\frac{1}{\sqrt{k!l!}}(a_{X}^{\dagger})^{k}(a_{Y}^{\dagger})^{l}|0,0\rangle$. 
We truncate the phonons at an $n$-phonon limit, where $k+l \leq  n$. 
We note that the two lowest eigenstates $\ket{\kappa_{1,2}}$ are always degenerate, thus they can be grouped as a degenerate level exhibiting $E$ symmetry unless the quadratic coupling becomes extremely strong~\cite{Bersuker_2006}, which is not the case of normally behaved defect centers.

We now turn to the description of the DFT calculations, in which the coordinates $X$ and $Y$ are treated as classical configurational variables entering the total energy of the full electron-nuclear system, rather than as quantized operators.
The total energy can be evaluated as a function of these distortion coordinates, thereby defining the adiabatic potential energy surface (APES).
The APES, which comprises two branches as illustrated in Fig.~\ref{fig:APES}, can be written as
\begin{equation}
    E_\pm = \dfrac{1}{2}\omega R^{2} \pm R \sqrt{ F^2+G^2R^2 + 2FGR \cos (3\phi) } \text{,}
\end{equation}
where $R=\sqrt{X^2+Y^2}$ is the distance from the conical intersection and $\phi=\arctan (Y/X)$ is the pseudo-rotation angle.
The resulting APES exhibits three types of critical points:
(i) the conical intersection at ${X} = 0$, ${Y} = 0$ where $\ket{e^{\text{DFT}}_{x,y}}$ are degenerate and exhibiting $E_\text{hs}$ energy. It is unstable due to trigonal symmetry–breaking distortions.
(ii) three equivalent minima, at which the system attains its lowest total energy $E_\text{min}$ due to trigonal symmetry–breaking distortion;
(iii) three equivalent saddle points exhibiting $E_\text{sp}$ total energy.
These special configurations can be identified through geometry optimizations performed with imposed symmetry constraints.
The procedure begins by initializing the atomic structure in its high-symmetry configuration -- e.g., $D_{3d}$ or $C_{3v}$ for group-IV–-vacancy centers or the NV, $\mathrm{N_3V}$, ($\mathrm{CH_3O}$) centers (radical), respectively -- and relaxing the atomic positions while preserving this high symmetry.
Next, the lattice is distorted along $+X$, lowering the symmetry to $C_{2h}$ or $C_{1h}$.
Further optimizing the atomic positions by DFT yields either the saddle-point energy $E_\text{sp}$ or the minimum point $E_\text{min}$; however, at this stage it is not known which of the two has been reached.
To identify the remaining configuration, one has to distort the opposite $-X$ additionally, which will identify the missing other special point.
We note that retaining a mirror-symmetry constraint during the relaxation provides a practical advantage: it prevents the system from leaving the $X$ axis and relaxing to $E_\text{min}$ beacause the system is confined in $Y=0$ thus ensuring both special points are being located unambiguously.

Our purpose of the DFT calculations is to parametrize the Jahn-Teller system. 
The full reconstruction of the APES makes possible to express the linear $F$ and quadratic $G$ Jahn-Teller coefficients as follows
\begin{equation}
	F = \sqrt{2E_{\text{JT}}\omega\left(1- \frac{\delta  }{2E_{\text{JT}}-\delta}\right)}  \qquad \text{and} \qquad  G = \frac{\omega\delta}{4E_{\text{JT}}-2\delta}\text{,} 
\end{equation}
where $E_{\text{JT}} = E_{\text{hs}}-E_{\text{min}} $ is the Jahn-Teller (JT) (stabilization) energy, $\delta = E_{\text{sp}}- E_{\text{min}}$ is the barrier height thus the quantities $E_{\text{JT}}$ and $\delta$ are readily obtained as total-energy differences directly from the DFT runs.
To determine $\omega$, we proceed as follows.
One may fit two one-dimensional harmonic oscillators along the $X$ coordinate (e.g. $Y=0$) in the neighborhood of configurations $E_{\text{min}}$ and $E_{\text{sp}}$.
Specifically, the two oscillators will exhibit
\begin{equation}
\omega_{\text{min}}=\!\!\!\sqrt{2\frac{E_{\text{hs}}-E_{\text{min}}}{d_{\text{min}}^{2}}}\!=\omega+2G\text{,}\;\;\omega_{\text{sp}}=\!\!\!\sqrt{2\frac{E_{\text{hs}}-E_{\text{sp}}}{d_{\text{sp}}^{2}}}\!=\omega-2G\text{,}
\end{equation}
frequencies where $d_{\text{min}}$ and $d_{\text{sp}}$ denote the distances between the high-symmetry configuration and the Jahn–Teller minimum, and between the high-symmetry configuration and the saddle-point geometry, respectively.
Both $d_{\text{min}}$ and $d_{\text{sp}}$ are generalized coordinates and can be expressed as follows:
\begin{equation}
	d_{\text{min/sp}}=\sqrt{\sum_{i}(\boldsymbol{d}_{\text{hs,}i}-\boldsymbol{d}_{\text{min/sp,}i})^{2}m_{i}} \text{,}
\end{equation}
where $\boldsymbol{d}_{\text{hs},i}$, $\boldsymbol{d}_{\text{min},i}$, $\boldsymbol{d}_{\text{sp},i}$ depicts the Cartesian coordinates of the $i$-th atom in the high-symmetry, minimum, or saddle-point configuration, respectively, and $m_{i}$ is the mass of the $i$-th nucleus.
And thus finally, $\omega$ of the two-dimensional oscillator can be expressed as the average of the two frequencies: $\omega=(\omega_{\text{sp}}+\omega_{\text{min}})/2$.

We now quantify the energy splitting induced by the spin–orbit interaction, $( \hat{H}_\text{SOC} =  \lambda_0\hat{L}_{z}\hat{S}_{z})$ acting within the subspace spanned by $\ket{\kappa_{1,2}}$. 
For sufficiently small $\lambda_0$, $\hat{H}_\text{SOC}$ may be treated as a perturbation.
By expanding $|\kappa_{1,2}\rangle$ in the real orbital basis $\{|e_x\rangle, |e_x\rangle \}$ and utilizing $\hat{L}_{z} =i  \left(\begin{smallmatrix}
		0 & -1\\
		1 & 0
	\end{smallmatrix}\right)$ one finds that the diagonal matrix elements vanish:
$\langle\kappa_{1}|\hat{L}_{z}|\kappa_{1}\rangle=\langle\kappa_{2}|\hat{L}_{z}|\kappa_{2}\rangle=0$. 
Only purely imaginary off-diagonal matrix elements remain, with magnitude of $|\langle\kappa_{1}|\hat{L}_{z}|\kappa_{2}\rangle|=p \times |\langle e_{x}|\hat{L}_{z}|e_{y}\rangle|$ 
where the $p$ Ham reduction factor~\cite{Ham_red_fact, Bersuker_2006} damps the $L_{z}$ angular moment by vibronic coupling.
Diagonalization of $\hat{H}_\text{SOC}$ in the two-dimensional subspace spanned by $|\kappa_{1,2}\rangle$ shows that the bare DFT spin–orbit coupling $\lambda_0$ is renormalized to an effective value:
\begin{equation}
\lambda^\text{pert.}_{\text{theory}}=p^\text{pert.}\times\lambda_0 \approx \lambda_{\text{expt}}
\text{.}
\label{eq:SOC_DJT}
\end{equation}

We note that this perturbative approach is no longer valid for heavier impurities, for which $\lambda_0$ is comparable or even exceeds $E_\text{JT}$.
Therefore, we emphasize the perturbative approach by the $\text{pert.}$ label for $\lambda^\text{pert.}_{\text{theory}}$ and $p^\text{pert.}$.
In such cases, the spin-orbit interaction can no longer be treated as a perturbation and must be included directly in the numerical diagonalization of the full Hamiltonian $\hat{H}$, as it can substantially affect and modify the vibronic expansion given in Eq.~\eqref{eq:waveansac}.

\subsection{$E\otimes e$ Jahn-Teller interaction and spin-orbit coupling}
\label{sec:SOCJT}
In this section, we take into consideration an additional degrees of freedom: the electron spin of the unpaired electron (or hole) that can affect the electronic orbital through spin-orbit coupling from~Eq.\eqref{Eq.TheSOC}.
Therefore, we extend the expansion of Eq.~\eqref{eq:waveansac} with electronic spin $\ket{\xi} = {\ket{\uparrow}, \ket{\downarrow}}$.
Therefore, we expand the eigenstates of the full $\hat{H} = \hat{H}_{\text{SOC}}+ \hat{H}_{\text{vib}} + \hat{H}_{\text{DJT}} $ Hamiltonian with the following expansion:
\begin{equation}
|\varepsilon_{i}\rangle=\!\sum_{\alpha}^{\{x,y\}}\sum_{\xi}^{\{\uparrow,\downarrow\}}\sum_{k,l}c_{k,l,\alpha,\xi}^{(i)}|e_{\alpha}\rangle\otimes|\xi\rangle\otimes|k,l\rangle
\text{,}
\label{eq:DJTSOC_waveexpansion}
\end{equation}
where $m_l = \{x$, $y\}$ labels depict the orbital quantum number while {$k, l$} are phonon occupation numbers. 
We note that in this step, the Hamiltonian terms are promoted to the full vibrational–electronic–spin Hilbert space e.g.: $\hat{H}_{\text{DJT}} \rightarrow \hat{H}_{\text{DJT}}\otimes\mathds{1}_{\text{spin}}$ (Eq.~\eqref{eq:H_DJT}), $\hat{H}_{\text{SOC}} \rightarrow \hat{H}_{\text{SOC}} \otimes \mathds{1}_{\text{vib}}$ (Eq.~\eqref{Eq.TheSOC}), $\hat{H}_{\text{vib}} \rightarrow  \hat{H}_{\text{vib}} \otimes  \mathds{1}_{\text{orb}} \otimes \mathds{1}_{\text{spin}}$ (Eq.~\eqref{eq:vib}) where $\mathds{1}_{\text{vib}}$, $\mathds{1}_{\text{orb}}$, spin $\mathds{1}_{\text{spin}}$ are identity operators for vibration, orbital, spin, respectively. 

The two bare electronic spin-orbit Kramers doublets $|{{E}_{\frac{1}{2}}}\rangle$, $|{{E}_{\frac{3}{2}}}\rangle$ introduced in Eq.~\eqref{eq:DFT_E12_E32} are transformed into $|{\tilde{E}_{\frac{1}{2}}}\rangle$, $|{\tilde{E}_{\frac{3}{2}}}\rangle$ due to the combined action of dynamic Jahn-Teller effect and spin-orbit coupling.
Therefore, the first four $|\varepsilon_{1-4}\rangle$ eigenstates that of Eq.~\eqref{eq:DJTSOC_waveexpansion} will correspond to the two polaronic $|{\tilde{E}_{\frac{1}{2}}}\rangle$, $|{\tilde{E}_{\frac{3}{2}}}\rangle$ doublets, whose energy separation is defined as
\begin{equation}
    \lambda_{\text{theory}} = \braket{\varepsilon_{3,4} |\hat{H} | \varepsilon_{3,4}} - \braket{\varepsilon_{1,2} |\hat{H}| \varepsilon_{1,2}} \approx \lambda_{\text{expt}} \text{,}
\end{equation}
 and thus it can be identified that $\lambda_0$ becomes renormalized to $\lambda_{\text{theory}}$ in the combined vibrational-electronic-spin system. 
 We note that $\lambda_{\text{theory}} = K_{\text{JT}} + \lambda_{\text{SOC}}$ can be decomposed into two parts: (i) the Jahn-Teller- and vibronic-induced term $K_{\text{JT}} = \langle{\varepsilon_{3,4} |\hat{H}_{\text{DJT}} + \hat{H}_{\text{vib}} | \varepsilon_{3,4}}\rangle - \langle{\varepsilon_{1,2} |H_{\text{DJT}} + \hat{H}_{\text{vib}}| \varepsilon_{1,2}}\rangle$ and (ii) the spin-orbit contribution $\lambda_{\text{SOC}} = \langle{\varepsilon_{3,4} |\hat{H}_{\text{SOC}}| \varepsilon_{3,4}}\rangle - \langle{\varepsilon_{1,2} |\hat{H}_{\text{SOC}}| \varepsilon_{1,2}}\rangle$. 
One can clearly recognize the non-perturbative character of the spin-orbit interaction: as $\lambda_0$ increases, the expansion coefficients $c_{k,l,\alpha,\xi}^{(i)}$ from Eq.~\eqref{eq:DJTSOC_waveexpansion}, associated with $|\varepsilon_{1,2}\rangle$ and $|\varepsilon_{3,4}\rangle$ will gradually diverge from one another.
Consequently, $K_{\text{JT}}$ becomes increasingly pronounced, and even the term $\lambda_{\text{SOC}}$ no longer the same as it was in Eq.~\eqref{eq:SOC_DJT} for example, in the case of the $\mathrm{SnV^-}$ and $\mathrm{PbV^-}$ vacancies.

\subsection{The case of external magnetic, electric fields and strain}
\label{sec:Bfield}
External magnetic field $\vec{B} = ( B_x,B_y,B_z )$ couples to both orbital ($L$) and spin ($S$) angular momentum of the electron via the Zeeman interaction:
\begin{equation}
    \label{eq:mag_int}
    \hat{H}_{\text{mag.}}=\mu_{\text{B}}g_{L}B_{z}\hat{L}_{z}+\mu_{\text{B}}g_{S}(B_{x}\hat{S}_{x}+B_{y}\hat{S}_{y}+B_{z}\hat{S}_{z})
    \text{,}
\end{equation}
where $\mu_{\text{B}} = 0.0579 \text{ meV/T}$ is the Bohr magneton, and $g_S = 2.0023$ and $g_L$ are the orbital and electron spin $g$-factor, respectively. 
We note that, here $g^{\text{gnd,ex}}_L$ factors are different values for optical ground (gnd) and excited (ex) states. 
Additionally, they differ from the $g_L=1$ value observed in full rotation group $\mathrm{SO(3)}$ because the defects we consider are exhibiting only trigonal 120$^{\circ}$ rotation symmetry thus  $g^{\text{gnd,ex}}_L$-s individual fractional values, see Appendix D in Ref.~\cite{PhysRevX2018} or Refs.~\cite{orb_red_fact, Gerloch_1968, Gajek_1985, Wei_2012} for details.
Thus, the Zeeman Hamiltonian $\hat{H}_{\text{mag.}}$ will lift the Kramers degeneracy for both $|{\varepsilon_{1,2}}\rangle$,  $|{\varepsilon_{3,4}}\rangle$ resulting in four states with distinct energies that can be seen in Fig.~\ref{fig:sketch} or \ref{fig:tin_ZPL}.
Therefore, $\hat{H}_{\text{mag.}}$ can be incorporated into the spin-orbit-phonon expansion that of Eq.~\eqref{eq:DJTSOC_waveexpansion} and then numerically determine the splitting due magnetic field.
We note that the vibrational degrees of freedom can be traced out~\cite{Ham_red_fact, PhysRev.166.307, Norambuena_2020} whenever one is only interested in the lowest four $|\varepsilon_{1-4}\rangle$ states. We will discuss it in the upcoming Section~\ref{sec:modelHamilton}.

The splitting induced by local strain or external electric field can be described by \cite{sil_vac, Meesala_2018, Doherty2013, PhysRevX.13.031022}
\begin{equation}
    \label{eq:strain_int}
\hat{H}_{\text{strain}}=-\varUpsilon_{x}\hat{\sigma}_{z}+\varUpsilon_{y}\hat{\sigma}_{x}
    \text{,}
\end{equation}
such that, in the presence of sufficiently strong strain or electric fields, the energy separation between the $|E_{\frac{3}{2}}\rangle$ and $|E_{\frac{1}{2}}\rangle$ Kramers doublets at zero magnetic field is enlarged to $\Delta > \ \lambda_{\text{theory}}$.
Since each specific defect center experiences a distinct local strain environment and a characteristic configuration of nearby charged defects (which can generate residual electric fields), the corresponding splitting, $\Delta$ is typically an emitter-specific parameter, measured individually for each single defect.
For inversion-symmetric centers such as G4V, the coupling to external electric fields vanishes. In contrast, systems lacking inversion symmetry -- such as the NV center -- exhibit finite electric-field coupling.
However, this coupling can be either an advantage or a disadvantage. (i) NV centers are prone to spectral diffusion~\cite{PhysRevApplied.18.064011, PhysRevLett.110.027401, PhysRevLett.116.033603}, in which the position of the zero-phonon line (ZPL) fluctuates over time due to charge-transfer processes involving nearby defects, thereby complicating their use in practical qubit applications. 
(ii) On the other hand, the same coupling is advantageous enhancing qubit's $T_2^*$ coherence time~\cite{PhysRevX2018, Jahnke_2015, PhysRevB.109.085414, fq19-lfmv, Brevoord_2025, PhysRevX.13.041037, Sohn_2018, Meesala_2018, PhysRevLett.128.153602, Kuruma_2024} by effectively quenching the decoherence induced by thermal phonons by enlarging $\Delta$ by means of strain engineering.
Here, we note that tracing over vibrational degrees of freedom is also convenient for strain: we will further discuss the $\Delta$ splitting in the following Section~\ref{sec:modelHamilton}.

\subsection{Effective model Hamiltonian}
\label{sec:modelHamilton}
The higher-lying vibronic states $|\varepsilon_{j>4}\rangle$ appearing in Eq.~\eqref{eq:DJTSOC_waveexpansion} are irrelevant for standard optical measurements, as the system relaxes into one of the four lowest-lying states $|\varepsilon_{1-4}\rangle$ on a timescale in the $\sim$ps to $\sim$fs regime.
This situation corresponds to the Kasha rule~\cite{Kasha_1950,Kasha_1956}, according to which only the vibronic ground levels are typically populated under optical excitation, unless the system is probed by ultrafast spectroscopy~\cite{Liu_2021, PhysRevB.97.220302, Huxter_2013, Ulbricht_2016, Carbery_2024, Eng_2015}, where the dynamics of these short-lived resonances can be characterized.
Thus, in typical state-of-the-art experiments, an effective four-level Hamiltonian ($\hat{H}_{\text{eff}}$) can be employed~\cite{Norambuena_2020, PhysRevX2018, PhysRevX.13.031022, PhysRevX.11.031021} to describe the $|\tilde{E}_{\frac{1}{2}}\rangle$, $|\tilde{E}_{\frac{3}{2}}\rangle$ manifold by introducing effective 
orbital: $\hat{L}_{z}^{\text{eff}}=\hat{\sigma}_{y}^{\text{eff}}=\bigl(\begin{smallmatrix}0 & -i\\ i & 0 \end{smallmatrix}\bigr)$, 
$\hat{\sigma}_{z}^{\text{eff}}=\bigl(\begin{smallmatrix}1 & 0\\ 0 & \!\!-1 \end{smallmatrix}\bigr)$, 
$\hat{\sigma}_{x}^{\text{eff}}=\bigl(\begin{smallmatrix}0 & 1\\ 1 & 0 \end{smallmatrix}\bigr)$
and spin:
$\hat{S}_{x}^{\text{eff}}=\frac{1}{2}\bigl(\begin{smallmatrix}0 & 1\\ 1 & 0 \end{smallmatrix}\bigr)$,
$\hat{S}_{y}^{\text{eff}}=\frac{1}{2}\bigl(\begin{smallmatrix}0 & -i\\ i & 0 \end{smallmatrix}\bigr)$,
$\hat{S}_{z}^{\text{eff}}=\frac{1}{2}\bigl(\begin{smallmatrix}1 & 0\\ 0 & \!\!-1 \end{smallmatrix}\bigr)$
operators:
\begin{equation}
\label{H_eff}
\begin{aligned}\underset{{\textstyle \Downarrow}}{\underbrace{\hat{H}}} & =\underset{{\textstyle \Downarrow}}{\underbrace{\hat{H}_{\text{SOC}}}}\quad+ & \cancel{\hat{H}_{\text{vib}}}+\cancel{\hat{H}_{\text{DJT}}}+\underset{{\textstyle \Downarrow}{\textstyle \Downarrow}}{\underbrace{\hat{H}_{\text{mag.}}}}+\underset{{\textstyle \Downarrow}}{\underbrace{\hat{H}_{\text{strain}}}}\\
\hat{H}_{\text{eff}} & =\lambda_{\text{theory}}\hat{L}_{z}^{\text{eff}}\hat{S}_{z}^{\text{eff}} & +\quad\mu_{\text{B}}\bigl(fB_{z}\hat{L}_{z}^{\text{eff}}+2\delta_{f}B_{z}\hat{S}_{z}^{\text{eff}}\bigr)\\
 &  & +\mu_{\text{B}}g_{S}\bigl(B_{x}\hat{S}_{x}^{\text{eff}}+B_{y}\hat{S}_{y}^{\text{eff}}+B_{z}\hat{S}_{z}^{\text{eff}}\bigr)\\
 &  & -\varUpsilon_{x}^{\text{eff}}\hat{\sigma}_{z}^{\text{eff}}+\varUpsilon_{y}^{\text{eff}}\hat{\sigma}_{x}^{\text{eff}}
\end{aligned}
\end{equation}

The first term represents the effective spin-orbit coupling ($\lambda_{\text{theory}}$).
The subsequent terms describe the interaction of the magnetic field with the effective orbital and spin angular momenta of the electron.
We note that $f = pg_L$ indicates that the orbital $g$-factor associated with $L_{z}$ is reduced by the Ham reduction factor $p$, since the orbital Zeeman term in Eq.~\eqref{eq:mag_int} contains the operator $\hat{L}_{z}$, which transforms according to the $A_2$ irreducible representation of the $C_{3v}$ point group.
Therefore, projection onto the effective subspace reduces the orbital angular momentum such that: $\langle\hat{L}_{z}\rangle_{\text{ph}}=p\hat{L}_{z}^{\text{eff}}$, where $\langle...\rangle_{\text{ph}}$ denotes averaging over the phonon bath (see Appendices A and B of Ref.~\cite{Norambuena_2020} for details).
In contrast, the spin Zeeman interaction remains unchanged in the effective model, as it involves only spin operators $\{\hat{S}_{x}$, $\hat{S}_{y}$, $\hat{S}_{z}\}$ and no orbital contribution.
However, an additional second-order term, $2\delta_{f}B_{z}\hat{S}_{z}^{\text{eff}}$ may arise for heavier G4V centers such as SnV$^-$ and PbV$^-$.
This contribution originates from the orbital Zeeman interaction but manifests as an effective spin operator $\hat{S}_{z}^{\text{eff}}$ in the effective system, see Appendix C in Refs.~\cite{PhysRevX2018, PhysRevX.10.039901}) in the reduced Hamiltonian (see Appendix C of Refs.~\cite{PhysRevX2018, PhysRevX.10.039901}).
The magnitude of this additional term can be determined utilizing the eigenstates $|\varepsilon_{1-4}\rangle$ defined in Eq.~\eqref{eq:DJTSOC_waveexpansion}:
\begin{equation}
\delta_{f}=g_{L}\times\bigl(\langle\varepsilon_{3,4}|\hat{L}_{z}\hat{S}_{z}|\varepsilon_{3,4}\rangle+\langle\varepsilon_{1,2}|\hat{L}_{z}\hat{S}_{z}|\varepsilon_{1,2}\rangle\bigr)
\text{.}
\end{equation}

Additionally, we incorporate the effect of external strain or transverse electric fields through the effective parameters $\varUpsilon_{x}^{\text{eff}}$ and $\varUpsilon_{y}^{\text{eff}}$. 
The bare strain Hamiltonian $\hat{H}_{\text{strain}}$ in Eq.~\eqref{eq:strain_int} is renormalized by the Ham factor $q=(1+p)/2$, because orbital operators $\{-\hat{\sigma}_{z}, \hat{\sigma}_{x}\}$ transform according to the $\{E_x, E_y\}$ irreducible representation of $C_{3v}$; see the discussion below Eq.~(11) in Ref.~\cite{Norambuena_2020}.
Consequently, in the effective description one obtains: $\{\langle\hat{\sigma}_{z}\rangle_{\text{ph}}=q\hat{\sigma}_{z}^{\text{eff}}$, $\langle\hat{\sigma}_{x}\rangle_{\text{ph}}=q\hat{\sigma}_{x}^{\text{eff}}\}$ which implies: $\{\varUpsilon_{x}^{\text{eff}}=q\varUpsilon_{x}$, $\varUpsilon_{y}^{\text{eff}}=q\varUpsilon_{y}\}$.
The resulting splitting between the two Kramers doublets at zero magnetic field is then enhanced by strain as we show below:
\begin{equation}
\Delta=\sqrt{\lambda_{\text{theory}}^{2}+4(\varUpsilon_{x}^{\text{eff}})^{2}+4(\varUpsilon_{y}^{\text{eff}})^{2}}
\text{.}
\end{equation}
We remark that the 4$\times$4 effective Hamiltonian $\hat{H}_{\text{eff}}$ reproduces the spectroscopic properties of the full Hamiltonian $\hat{H}$ in Eq.~\eqref{eq:gen_H}, while avoiding the excessive computational cost associated with explicitly treating the large number of phononic degrees of freedom.
Thus, for most practical purposes, the effective model provides a sufficient and efficient description.

\section{Numerical results on solid state quantum bits in diamond}
\label{sec:numericalDFT}
\begin{figure}[h]
\includegraphics[width=0.35\textwidth]{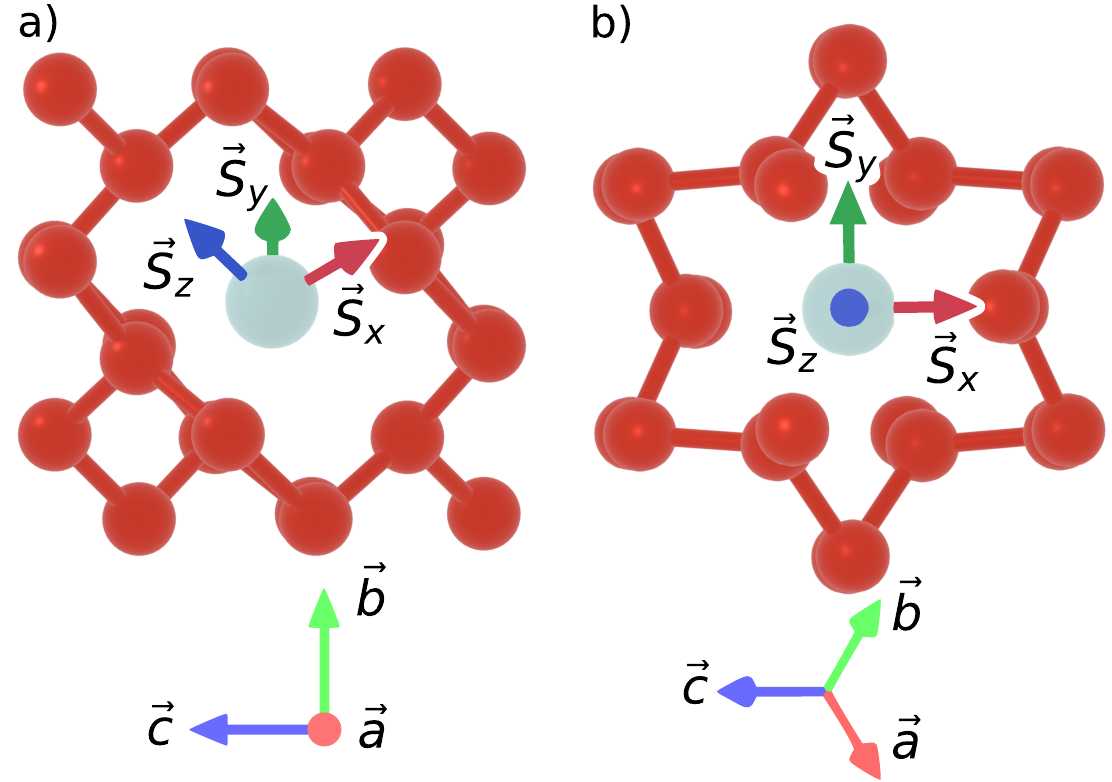}
\centering
\caption{
This figure illustrates a group-IV vacancy defect embedded in a diamond crystal. 
The crystallographic basis vectors ($\vec{a}$, $\vec{b}$, $\vec{c}$) are shown at the bottom of the figure in light red, green, and blue, respectively. 
The basis vectors for defect’s spin ($\vec{S}_x$, $\vec{S}_y$, $\vec{S}_z$) are depicted in darker shades and needs to be defined in the configuration file.
For G4V centers, the spin $z$-axis aligns with the [111] crystallographic direction, which corresponds to the $C_{3}$ rotation axis. The two  $x$- and $y$-axes can be freely chosen freely. 
In panel (a), the [100] crystallographic direction points toward the viewer, while in panel (b), the defect’s $C_3$ rotational axis points toward the viewer.
}
\label{fig:3D_point_defect}
\end{figure}

In this section, we present an illustrative example fine-structure calculation for G4V centers in diamond, a class of well-studied point defects shown in Fig.~\ref{fig:3D_point_defect}.
As a specific example, we also analyze the ZPL splitting of the $\text{SnV}^-$ center in an external magnetic field.
The electronic structures of the high-symmetry (hs), minimum energy (min), and saddle-point (sp) geometries, as well as the spin–orbit coupling, can be evaluated by means of state-of-the-art DFT methodology.
The $g$-factor $g_L$, however, cannot be calculated by $\mathtt{VASP}$ faithfully.
In certain special cases -- specifically, when the relevant state is strongly localized on the dopant’s $d$-orbitals such that the entire orbital moment is confined to a single atom and contained within its PAW sphere~\cite{PAW_method, PhysRevB.59.1758, PhysRevB.62.11556} -- $g_L$ can be extracted directly via the $\mathtt{LORBMOM}$ tag~\cite{Csore2020}.
For the SiV center in diamond, however, the orbital moment is distributed over six carbon dangling bonds (see Fig. 8 in Ref.~\cite{PhysRevX2018}), which renders this approach inapplicable.
A more sophisticated treatment -- such as those implemented in advanced quantum-chemistry packages~\cite{Neese_2005, Bruder_2022, Franzke_2022} -- would be required; however, these methods can model solids only through very small finite clusters, introducing severe finite-size artifacts.
In that case, the ground- and excited-state values of $g_L$ were obtained by fitting the effective Hamiltonian to experimental data, yielding $g_L^\text{gnd}=0.328$ and $g_L^\text{exc}=0.782$, respectively. 
These values are commonly assumed to be transferable to the other G4V centers~\cite{PhysRevX2018}.

It is important to emphasize that the natural basis of the defect spin system is, in general, not aligned with the crystallographic basis vectors. 
In the present example, we employ a simple-cubic supercell of diamond, where the supercell lattice vectors coincide with those of the Bravais cell.
G4V centers exhibit a high-symmetry axis -- corresponding to the $C_3$ rotation axis -- aligned with the [111] crystallographic direction of diamond.
It is therefore natural to adopt the [111] direction as the spin quantization axis, i.e., the $z$-axis axis associated with the electronic spin operator $\hat{S}_z$ appearing in Eq.~\eqref{eq:SOC_DJT}.
The remaining basis vectors $x$- and $y$-axes may be chosen arbitrarily, provided they are orthogonal to the $z$-axis and form a right-handed coordinate system for convenience.
$\mathtt{Exe.py}$ requires a .cfg configuration file as input to specify the following attributes of the system in order to calculate the $E\otimes e $ Jahn-Teller interaction:
\begin{itemize}
  \item vasprun.xml outputs of $\mathtt{VASP}$ DFT calculations with relaxed atomic positions at $E_\text{hs}$, $E_\text{min}$, $E_\text{sp}$ special points, see Fig.~\ref{fig:APES} 
  \item maximum number of phonon quanta ($k+l\leq n$) in Eqs.~\eqref{eq:DJTSOC_waveexpansion},~\eqref{eq:waveansac}. Fig. \ref{fig:quanta_conv} shows how the $p$ factor converges with respect to this parameter. In practice we set it to 12.
  \item spin-orbit coupling $(\lambda_0)$ as described in section~\ref{sec:SOC}
\end{itemize}
For calculating the ZPL fine structure in the presence of magnetic field, it is further necessary to specify the following parameters:
\begin{itemize}
  \item orbital reduction factor $g_L$
  \item the basis vectors for electronic spin quantization
  \item range and directions of the applied magnetic field \\in terms of the basis vectors of geometrical configurations
\end{itemize}

The code automatically normalizes the user-specified basis vectors.
The external magnetic field is then transformed into the coordinate system defined by the defect’s spin basis. The implementation is explicitly interfaced with the $\mathtt{VASP}$ package~\cite{vasp}.
Nevertheless, the code is also compatible with results from other electronic-structure calculations, since the relevant energies can be provided directly via the configuration file and the geometries can be supplied in .csv format. 
In this latter case, the user must additionally specify the following attributes in the .cfg file:
\begin{itemize}
    \item energies of the geometries $(E_{\text{hs}}, E_{\text{min}}, E_{\text{sp}})$
    \item basis vectors of the geometry
    \item atomic masses for ions
\end{itemize}

Moreover, the input may be specified either in terms of the Jahn-Teller parameters ($E_{\text{JT}}, \delta, d_{\text{min}}, d_{\text{sp}}$) or via the Taylor-expansion coefficients $(F, G)$ of $\hat{H}_{\text{DJT}}$ as defined in (\ref{eq:H_DJT}).
All supported use cases are documented in the $\mathtt{point\_defect\_config\_files}$ directory of the GitHub repository. 
The raw parameters extracted from the input files are summarized in Table~\ref{tab:raw_pars}.
The code generates the following output files:
\begin{itemize}
    \item eigenvalues and eigenstates in separete .csv files
    \item all raw parameters and theoretical values that are calculated, describes in Sec. \ref{sec:theory}
    \item energies of A,B,C,D transitions
    \item expectation values of $L_z\otimes S_z$ \text{.}
\end{itemize}

\begin{table*}[h]
    \centering
    \caption{Raw parameters extracted from the vasprun.xml $\mathtt{VASP}$ calculations supplemented with additional data for $g_L$ and $\lambda_0$ in an accompanying configuration file. Those are the energies of the high symmetric, global energy minimum and saddle point geometries $(E_{\text{hs}}, E_{\text{min}}, E_{\text{sp}})$ and the distances between them $(d_{\text{min}}, d_{\text{sp}})$. From these values the parameters of the Jahn-Teller interaction are calculated. $\lambda_0$ is the energy splitting between degenerate orbitals (Kohn-Sham levels) caused by spin-orbit coupling calculated using DFT. The orbital reduction factor $(g_L)$ appears in the magnetic field - angular momentum interaction in Eq.~\eqref{eq:mag_int}. }
    \label{tab:raw_pars}
    \begin{tabular}{llllllll}
    \hline 
        defect type 
        & \makecell{$E_{\text{hs}}$\\(eV)} 
        & \makecell{$E_{\text{min}}$\\(eV)} 
        & \makecell{$E_{\text{sp}}$\\(meV)} 
        & \makecell{$d_{\text{min}}$ \\ 
        $(\AA\!\ensuremath{\sqrt{\mathrm{amu}}}$) }
        & \makecell{$d_{\text{sp}}$ \\
        $(\AA\!\ensuremath{\sqrt{\mathrm{amu}}}$) }
        & $g_{L}$
        & \makecell{ $\lambda_0$ \\ (meV) } \\ \hline
        SiV gnd & -5375.9368 & -5375.9791 & -5375.9763 & 0.2085 & 0.2109 & 0.328 & 0.8 \\ 
        SiV ex & -5374.1818 & -5374.2603 & -5374.2577 & 0.3442 & 0.3450 & 0.782 & 6.9 \\ 
        GeV gnd & -5372.7556 & -5372.7858 & -5372.7837 & 0.1854 & 0.1884 & 0.328 & 2.2 \\ 
        GeV ex & -5370.5590 & -5370.6440 & -5370.6390 & 0.3574 & 0.3530 & 0.782 & 36.1 \\ 
        SnV gnd & -5368.3068 & -5368.3284 & -5368.3268 & 0.1644 & 0.1676 & 0.328 & 8.6 \\ 
        SnV ex & -5366.1366 & -5366.2197 & -5366.2129 & 0.3407 & 0.3421 & 0.782 & 95.9 \\
        PbV gnd & -5364.4629 & -5364.4785 & -5364.4779 & 0.1538 & 0.1483 & 0.328 & 34.6 \\ 
        PbV ex & -5361.9405 & -5362.0324 & -5362.0198 & 0.3436 & 0.3362 & 0.782 & 245.2 \\ \hline
    \end{tabular}
\end{table*}

\subsection{Computational details}
\label{sec:compdet}
The accuracy of the results obtained with $\mathtt{jahn {\text -} teller {\text -} dynamics}$ is profoundly impacted by the precision of input parameters.
In this work, we employ density functional theory (DFT) within the Born-Oppenheimer approximation, as implemented in the $\mathtt{VASP}$ 5.4.1 code~\cite{vasp} within plane-wave supercell framework and projector-augmented-wave (PAW) formalism~\cite{PAW_method,PAW_method2}.
We note that our code is also compatible with newer 6.x.x versions.
However, we observed that numerical stability of constrained occupation $\Delta$SCF calculations by setting $\mathtt{FERWE}$, $\mathtt{FERDO}$ tags often vary from version to version see Refs.~\cite{PhysRevApplied.22.034056, xiong2025deltascftextttvaspexcitedstatedefect} for additional details reported by other research groups.
Therefore, we opted to $\mathtt{VASP}$ 5.4.1 where we found the best convergence over the years.
Calculations are performed utilizing the Perdew-Burke-Ernzerhof (PBE)~\cite{PBE} generalized-gradient functional or the Heyd-Scuseria-Ernzerhof~\cite{HSE_1, HSE_2} (HSE06) hybrid functional. 
Point defects are modeled with an plane-wave energy cutoff of 370 eV, and strict convergence criteria on forces action on ions: $10^{-4} \text{eV/\AA}$ during the geometry optimization.
We used 512-atom supercells to mitigate finite size effects and obtain results suitable for comparison with experiment except the spin-orbit parameters that we discuss next.

We employed the noncollinear formalism implemented in $\mathtt{VASP}$ 5.4.1~\cite{PhysRevB.62.11556, PhysRevB.93.224425} to determine the spin-orbit ($\lambda_0$) parameters.
We note that fully converged spin–orbit splittings generally require supercells larger than the 512-atom cell mentioned before.
The spin quantization axis was chosen along the [111] direction ($C_3$ rotation axis).
We determine the spin-orbit parameters in supercells optimized in optimized geometries previously by means of conventional collinear calculations; thus, the influence of spin–orbit coupling on the ionic positions was neglected.
The occupations of the relevant degenerate Kohn–Sham orbitals in the band gap were constrained to an equal (half–half) distribution, corresponding to a single-electron configuration $(e_{+\uparrow}^{1}e_{-\uparrow}^{1}e_{+\downarrow}^{0.5}e_{-\downarrow}^{0.5})$.

\subsection{Optical transition energies under magnetic fields}
In this section we provide exemplary calculations with the negatively charged group-IV--vacancy (G4V) defects in diamond.
In these defects, the impurity atom (Si, Ge, Sn, or Pb) adopts the split-interstitial configuration (see Fig.~\ref{fig:APES}(c)) exhibiting $D_{3d}$ point group symmetry.
G4V centers exhibits a $(e_{u}^{4}e_{g}^{3})$ occupation, leaving a single hole in the degenerate $e_g$ \textit{gerade} orbital predominantly localized on carbon dangling bonds surrounding the two vacancies~\cite{PhysRevX.10.039901}.
Their optically excited state can be characterized by $(e_{u}^{3}e_{g}^{4})$, thus, the hole is now present in the degenerate $e_u$ \textit{ungerade} orbital. 
Consequently, both the ground and excited states are subjects of the $E\otimes e$ Jahn-Teller problem~\cite{PhysRevX2018}, and the corresponding spin-orbit coupling enters with negative sign, $-\lambda_0$ according to Hund's third law, that fact also visible in $\mathtt{VASP}$ calculations.

\begin{table*}[!ht]
    \centering
    \caption{Jahn-Teller parameters and reduction factors that describe the $E\otimes e$ system, $\lambda_0$ is the spin-orbit splitting calculated by DFT, $E_{\text{JT}}$ is the Jahn-Teller energy, $\delta$ is the barrier energy, $\omega$ is the vibration energy quantum, $p$ is the Ham reduction factor, $g_L$ is the orbital reduction factor, $\lambda_{\text{theory}}$ is the theoretical spin-orbit splitting energy, $\lambda_{\text{expt}}$ is the experimental value taken from \cite{sil_vac},\cite{GeV_exp2015},\cite{tin-vac_diamond},\cite{PbV_exp}. 
    The table also contains $f$, $\delta_f$ factors that are necessary in order to configure the four state model described by Eq.~\eqref{H_eff}. 
    We note that $p$ and $\lambda_{\text{theory}}$ that of Eq.~\eqref{H_eff} slightly differs than the perturbative approach of $\lambda^\text{pert.}_{\text{theory}}$ and $p^\text{pert.}$ that was in Eq.~\eqref{eq:SOC_DJT}.
    The difference negligible for SiV, GeV, but becomes non-negligible for SnV, PbV please see the last four columns of Table. III. in Ref.~\cite{PhysRevX.10.039901}, where the last four $\{p$, ${\lambda_{\text{Ham}}}$,
    ${\lambda}$,
    ${\lambda_{\text{expt}}}\}$ columns depicts the $\{p^{\text{pert.}}$, ${\lambda_{\text{theory}}^{\text{pert.}}}$,
    ${\lambda_{\text{theory}}}$, ${\lambda_{\text{expt}}}\}$ parameters in the present article.
    }
    \label{tab:JT_pars}
    \begin{tabular}{llllllllll}
    \hline
        defect type & \makecell{ $E_{\text{JT}}$ \\(meV)} & \makecell{ $\delta$ \\(meV)} & \makecell{$\omega$\\ (meV)} & \makecell{$p$}  &\makecell{ $\lambda_0$ \\(meV)} & \makecell{ $\lambda_{\text{theory}}$ \\(meV)} & \makecell{ $\lambda_{\text{expt}}$ \\(meV)}& \makecell{$f$} & \makecell{$\delta_f$} \\ \hline
        SiV gnd & 42.3 & 2.9 & 88.2 & 0.316 & 0.8 & 0.25 & $0.21^{\text{\cite{sil_vac}}}$ & 0.104  & 0.001  \\ 
        SiV ex & 78.5 & 2.7 & 73.7 & 0.128 & 6.9 & 0.90 & $1.08^{\text{\cite{sil_vac}}}$ & 0.100 & 0.018\\ 
        GeV gnd & 30.1 & 2.1 & 83.4 & 0.394 & 2.2 & 0.87 & $0.75^{\text{\cite{GeV_exp2015}}}$ & 0.129 & 0.004 \\ 
        GeV ex & 85.0 & 5.0 & 73.9 & 0.117 & 36.1 & 4.21 & $4.63^{\text{\cite{GeV_exp2015}}}$ & 0.091 & 0.089 \\ 
        SnV gnd & 21.6 & 1.6 & 79.5 & 0.471 & 8.3 & 3.92 & $3.52^{\text{\cite{tin-vac_diamond}}}$ & 0.154 & 0.015 \\ 
        SnV ex & 83.2 & 6.8 & 75.6 & 0.125 & 95.9 & 12.10 & $12.41^{\text{\cite{tin-vac_diamond}}}$ & 0.098 & 0.232 \\ 
        PbV gnd & 15.6 & 0.6 & 74.9 & 0.494 & 34.6 & 18.14 & $17.5^{\text{\cite{PbV_exp}}}$ & 0.162 &0.069  \\ 
        PbV ex & 91.9 & 12.6 & 78.6 & 0.105& 245.2 & 28.57 & ~&0.082 &0.472 \\ \hline
    \end{tabular}
\end{table*}

\begin{figure}[h]
\includegraphics[width=0.45\textwidth]{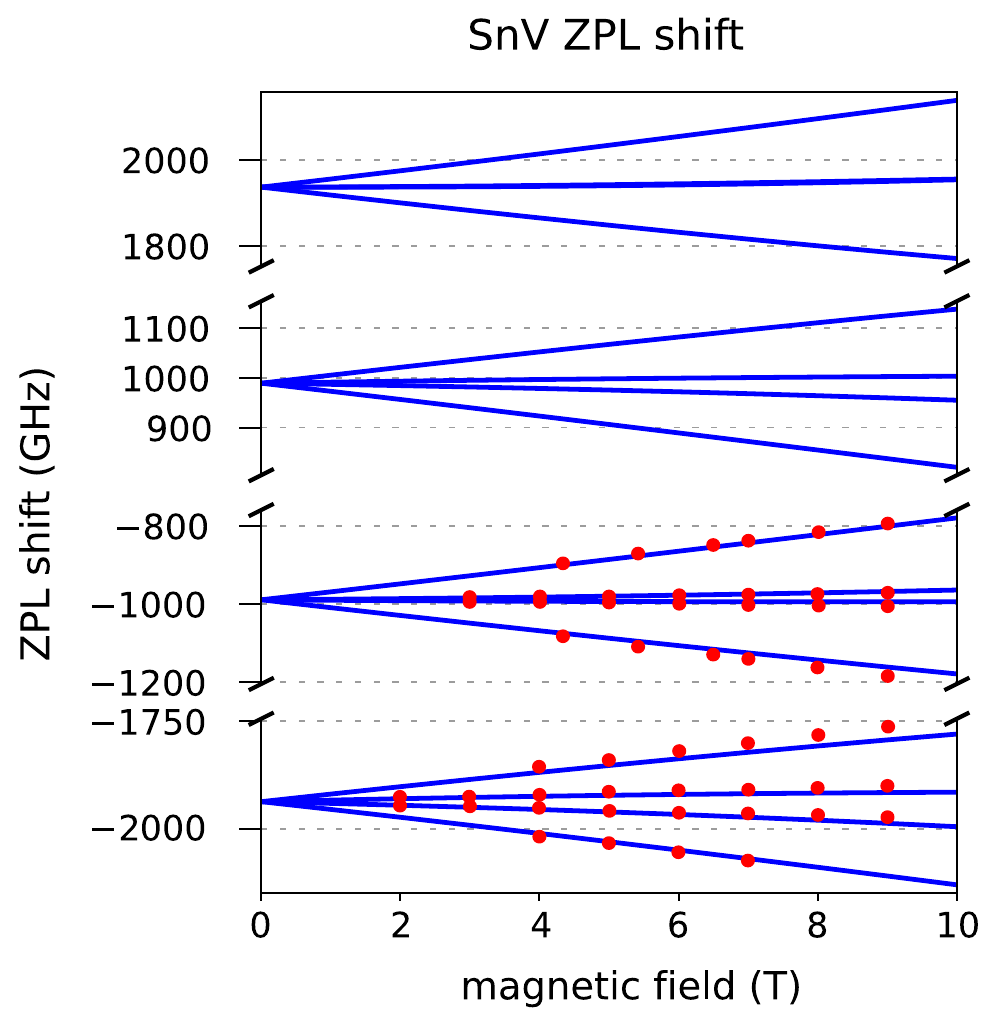}
\centering
\caption{Magnetic field strength dependence of ZPL fine strucure that of the $\text{SnV}^-$ defect in diamond. Experimental data points are taken from Ref.~\cite{tin-vac_diamond} are compared to the result of our $\mathtt{Exe.py}$ code for the negatively charged $\text{SnV}^-$ defect in diamond.}
\label{fig:tin_ZPL}
\end{figure}

In order to simulate the ZPL fine structure, parameters for both ground and excited states must be specified along with the applied magnetic field.
In the examples considered here, the magnetic field is aligned with the defect spin, i.e., along the [111] crystallographic direction.
The script $\mathtt{Exe.py}$ extracts the total energies of the geometries $(E_{\text{hs}}$, $E_{\text{min}}$, $E_{\text{sp}})$, the distances between them $(d_{\text{hs}}$, $d_{\text{min}}$, $d_{\text{sp}})$, orbital $g$-factors, $g_L$, and spin-orbit $\lambda_0$ parameters, are summarized in Table~\ref{tab:raw_pars} as obtained from DFT.
The code iterates over the user-specified magnetic-field range and constructs the corresponding Hamiltonian for each field value. 
It then computes the eigenenergies ($|\varepsilon_i\rangle$-s) for both the ground and excited states and subsequently determines the resulting ZPL fine structure, which can be directly compared with experimental observations, see Fig.~\ref{fig:tin_ZPL}.

At zero magnetic field, four optically allowed transitions (A, B, C, D) connect the fourfold-degenerate ($e_g$, $e_u$) manifolds (Fig.~\ref{fig:sketch}). Under nonzero magnetic field, each of these transitions acquires a distinct energy, giving rise to a characteristic fine structure in the spectrum with a total 16 individual transitions. 
In addition, the code determines the Jahn-Teller parameters and the theoretical spin-orbit splitting $\lambda_{\text{theory}}$ that is comparable with experimental data $\lambda_{\text{expt}}$ as shown in Table~\ref{tab:JT_pars}.

\begin{figure}[h!]
\includegraphics[width=0.45\textwidth]{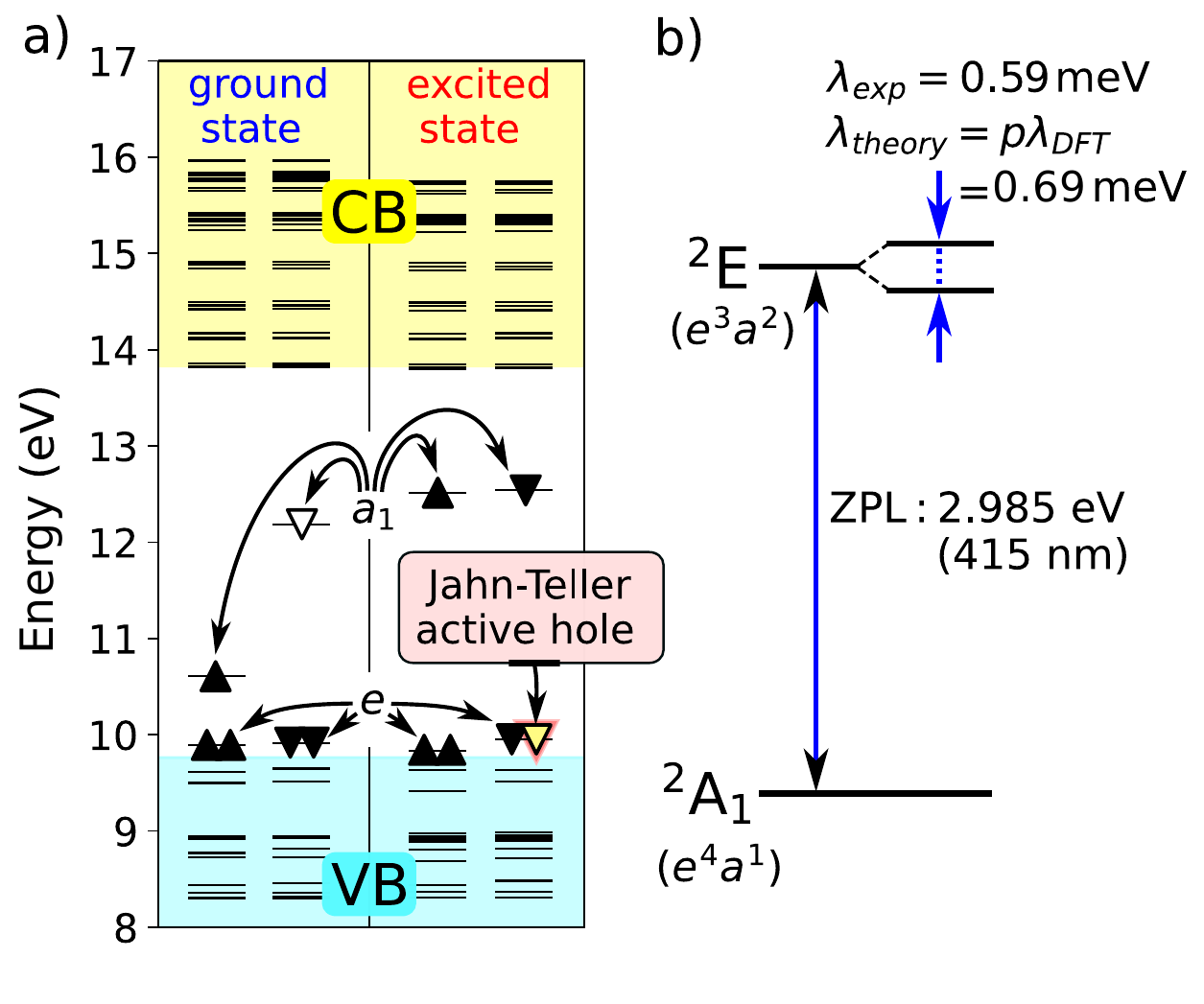}
\centering
\caption{In the ground state configuration, $e$ electron state is fully occupied. As we excite the system, the electron is promoted to the $a_1$ state and leaves a hole behind in the $e$ state. This leads to a dynamic Jahn-Teller system in the optical excited state.}
\label{fig:el_str}
\end{figure}

\begin{figure}[h!]
\includegraphics[width=0.35\textwidth]{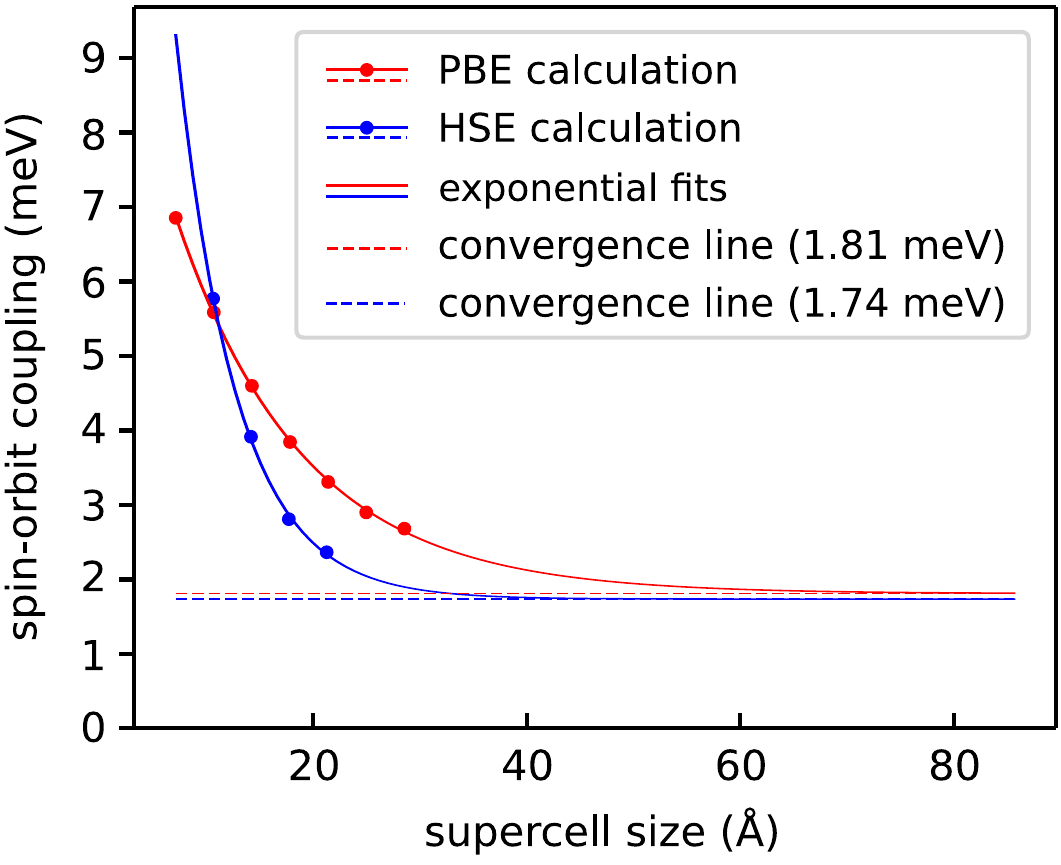}
\centering
\caption{ Spin-orbit $\lambda_0$ parameters using PBE (HSE06) functional in diamond supercells with superlattices ranging 63-4095 (63-1727) atoms. Both functionals show an exponential decay with respect to supercell length converging to $1.81$~meV ($1.74$~meV)}
\label{fig:SOC}
\end{figure}

\begin{figure}[h!]
\includegraphics[width=0.40\textwidth]{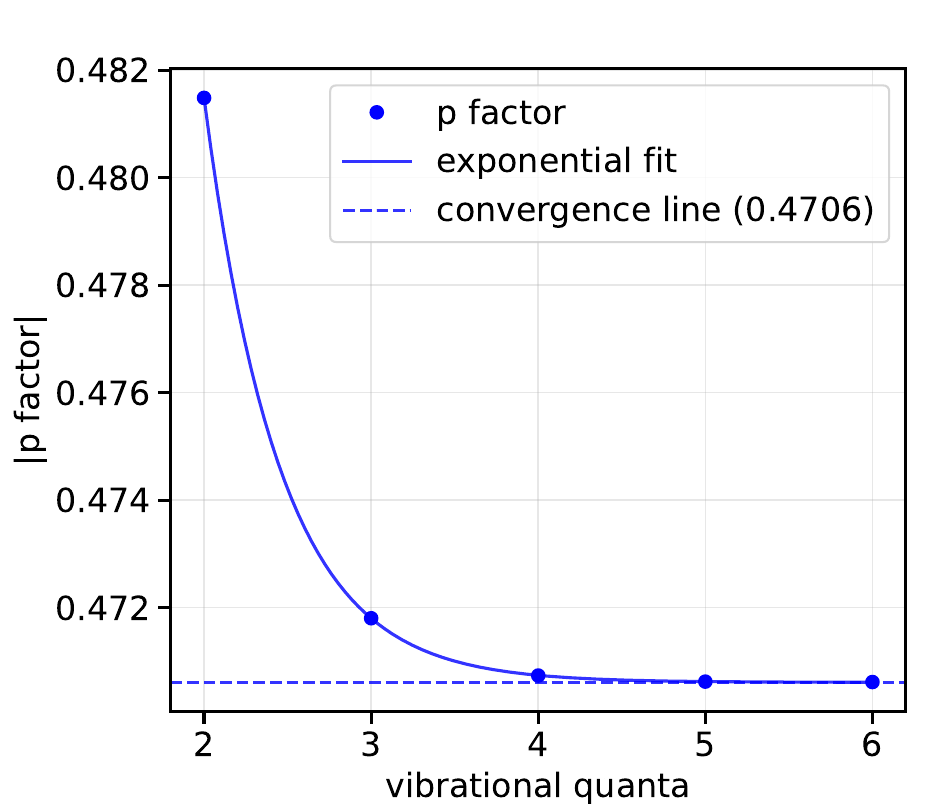}
\centering
\caption{ This figure shows the convergence of the $\text{SnV}^-$ center's $p$ factor with respect to the maximum number of vibrational quanta in each direction $\hat{X}, \hat{Y}$. It goes to $p = 0.4706$ after having at least 6 maximum phonon quanta in the system.}
\label{fig:quanta_conv}
\end{figure}

\begin{figure}[h!]
\includegraphics[width=0.45\textwidth]{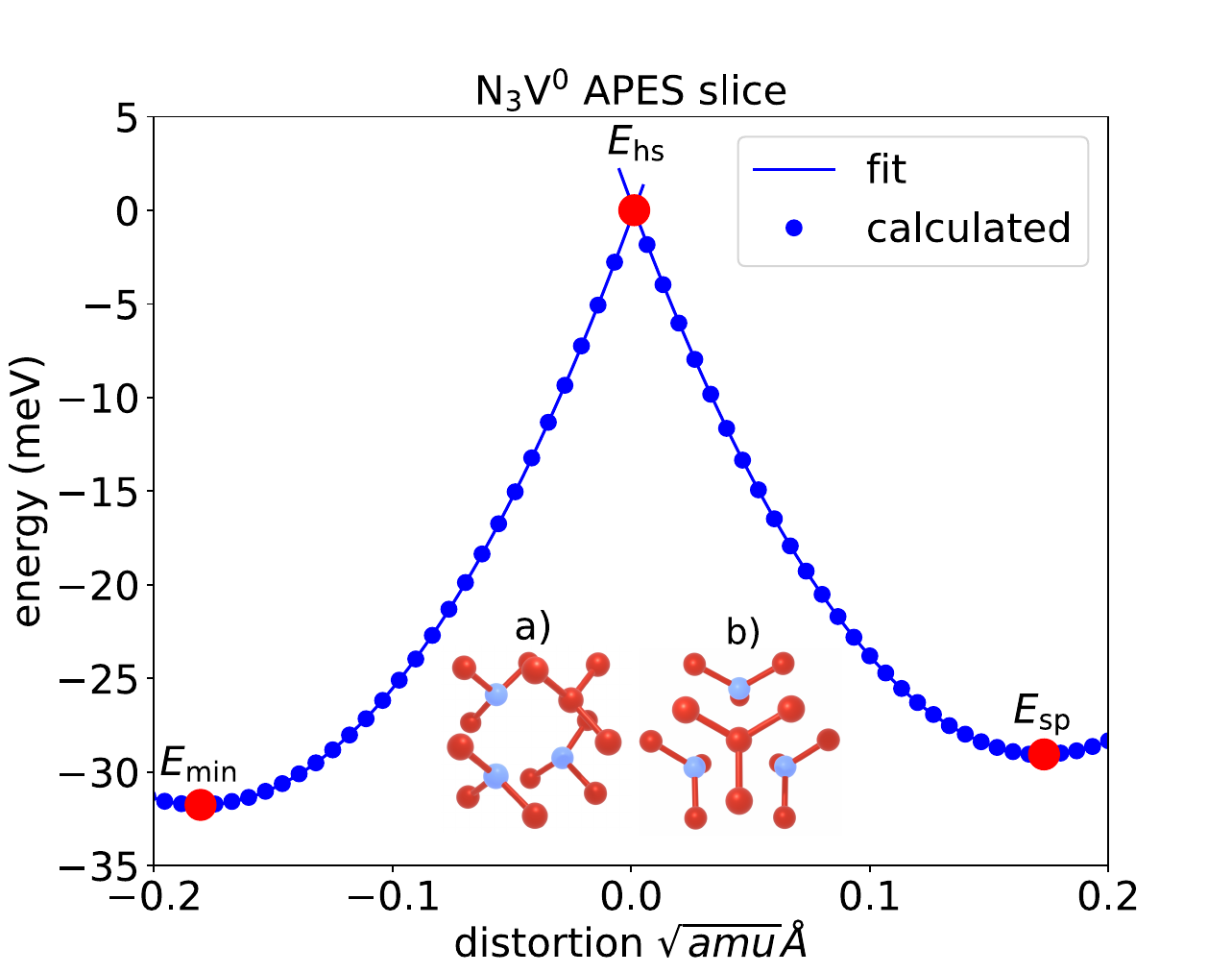}
\centering
\caption{This figure shows the calculated APES of the excited state of the neutral $ \text{N}_{3}\text{V}$ defect. The a) and b) part of the figure represents the $\text{N}_3\text{V}^0$ point defect.}
\label{fig:N3V_PES}
\end{figure}

\subsection{$\mathrm{N_3V^0}$ defect in diamond}
\label{sec:N3V}
In this section, we demonstrate the predictive capability of our method by applying $\mathtt{Exe.py}$ on the neutral $\text{N}_3\text{V}^0$ defect in diamond.
$\text{N}_3\text{V}^0$ center is made of three substitutional nitrogen atoms surrounding a vacancy embedded in the diamond host exhibiting $C_{3v}$ symmetry \cite{Ashfold2020}.
This defect gives rise to a prominent optical feature in most type-Ia natural diamonds containing B-type nitrogen aggregates and can also be generated artificially via nitrogen ion implantation followed by annealing at temperatures above 1200~$^\circ$C \cite{Zaitsev_2001}.
Figure~\ref{fig:el_str} schematically illustrates the electronic structure of the defect: $a_1$ and $e$ levels appear within the boundaries of the band gap, with the corresponding orbitals predominantly localized on carbon and nitrogen dangling bonds, respectively.
In the ground state configuration ($e^4a^1$), the $e$ level is fully occupied, therefore, it is not Jahn-Teller active $|^2A_1\rangle$ multiplet.
Upon optical excitation, an electron is promoted from the $e$ level to the $a_1$ level, leaving a hole in the $e$ level~\cite{BABAMORADI201717} and thus, the excited state is Jahn-Teller active $|^2E\rangle$ multiplett.
Experimentally, the ZPL of $\mathrm{N_3V^0}$ has been reported at 2.985 eV, with a fine-structure splitting of 0.59 meV in photoluminescence (PL) spectra~\cite{Zaitsev_2001}.
Utilizing the HSE06 functional, we obtain a ZPL energy as 3.059 eV, representing an improvement over an earlier (semi)local DFT result of 2.8 eV~\cite{Goss}.
We attribute the observed ZPL splitting to spin-orbit splitting within the $^2E$ excited state; thus, we employ our $\mathtt{Exe.py}$ code to calculate $\lambda_{\text{theory}}$ for the PL spectrum.
To generate all input parameters for $\mathtt{Exe.py}$, we performed constrained DFT calculations by means of the HSE06 functional on an 1728-atom supercell that holds 6911 valence electrons. 
In order to simulate the excited state (Fig. \ref{fig:el_str}) we employ constrained occupation DFT calculations with $\mathtt{FERWE = 3456*1.0}$ and $\mathtt{FERDO = 3454*1.0\:\:0.0\:\:1.0}$ tags.
These points perfectly fit two quadratic polynomials that intersect exactly at the $C_{3v}$ high-symmetry configuration, see Fig.~\ref{fig:N3V_PES}.
Finally, the derived Jahn-Teller energy of the system $E_{\text{JT}} = 32.12$~meV, the barrier energy $\delta = 2.68$~meV, and the vibrational energy quantum $\omega = 89.54$~meV as yielded by $\mathtt{Exe.py}$ code which results in a Ham reduction factor $p = 0.40$ for this system.

Next, we determine the bare spin-orbit $\lambda_0$ parameter for the $^2E$ excited state, where we force the $(e_{+\uparrow}^{1}e_{-\uparrow}^{1}e_{+\downarrow}^{0.5}e_{-\downarrow}^{0.5})$ occupation -- for example on the 1727-atom supercell -- by 
\begin{equation}
\mathtt{FERWE = 3455*1.0\:\:3453*1.0 \:\:0.5\:\: 0.5\:\: 2*1.0\:\: 0.0\:\:\ldots}    
\end{equation}
where Kohn-Sham energy difference between $e_{+\downarrow}^{0.5}$ and $e_{-\downarrow}^{0.5}$ single particle orbitals yields $\lambda_0$. The straightforward approach would be calculating the system in both
orbital configurations $(e_{+\uparrow}^{1}e_{-\uparrow}^{1}e_{+\downarrow}^{1}e_{-\downarrow}^{0})$
and $(e_{+\uparrow}^{1}e_{-\uparrow}^{1}e_{+\downarrow}^{0}e_{-\downarrow}^{1})$ with integer occupations
and identify their total energy difference as the $\lambda_0$.
However, in our experience this approach is not only computationally
more demanding. 
Unlike molecular systems with limited degrees of freedom, defects in solids experience more severe 
More importantly, the latter case $(e_{+\uparrow}^{1}e_{-\uparrow}^{1}e_{+\downarrow}^{0}e_{-\downarrow}^{1})$
can often be numerically unstable because it is not a ground state configuration, and thus it may relax to  $(e_{+\uparrow}^{1}e_{-\uparrow}^{1}e_{+\downarrow}^{1}e_{-\downarrow}^{0})$ or even quench spin-orbit coupling by converging into a real valued solution $(e_{x\uparrow}^{1}e_{y\uparrow}^{1}e_{x\downarrow}^{1}e_{y\downarrow}^{0})$.
Therefore, it is more convenient and reliable
to depend on the $(e_{+\uparrow}^{1}e_{-\uparrow}^{1}e_{+\downarrow}^{0.5}e_{-\downarrow}^{0.5})$
occupation even when the two electron halves effectively repel each
other, thus one would obtain a significantly larger total energy ($\sim0.2$
eV) in the fractional case by means of hybrid functionals such as
HSE06. 
We note that semilocal GGA (generalized-gradient approximation)
functionals such as PBE this on-site interaction seemingly does not
exist thus the fractional occupation case for PBE exhibits exactly the same total energy
as the two integer occupation cases, of course, except for the
small difference due to $\lambda_0$. Nevertheless, we
have shown in our previous~\cite{Thiering2017} work that the Kohn-Sham
energy difference can be interpreted as $\lambda_0$ because
it is numerically exactly the same as the aforementioned total energy
difference between the $(e_{+\uparrow}^{1}e_{-\uparrow}^{1}e_{+\downarrow}^{1}e_{-\downarrow}^{0})$
and $(e_{+\uparrow}^{1}e_{-\uparrow}^{1}e_{+\downarrow}^{0}e_{-\downarrow}^{1})$
occupations.

We used a sequence of diamond supercells of increasing size 63, 215, 511, 999, 1727, 2743, 4095 atoms to fully avoid the spurious interaction between periodic images.
Figure~\ref{fig:SOC} shows the calculated spin-orbit energies $\lambda_0$ as a function of supercell size; the values decrease exponentially with increasing superlattice length and converge to 1.81 meV (1.74 meV) within PBE (HSE06).

Combining this with the Ham reduction factor $p=0.40$, we obtain an effective spin-orbit splitting $\lambda_{\text{theory}}=0.69$~meV which is a very good agreement against the experimental data at $\lambda_{\text{expt}}=0.59$~meV, and thus, reaffirms N$_3$V$^0$ assignment for the 3.059 eV color center in diamond.

\subsection{$\mathrm{CH}_3\mathrm{O}^0$ radical in the single mode case}
\label{sec:CH3O}
 In this section we present how our code can be used to calculate the spin-orbit coupling in the $\text{CH}_3\text{O}$ radical.
 This radical possess $\text{C}_{3v}$ symmetry.
 We performed density functional theory calculations using the PBE0 functional \cite{PBE0} which combines the generalized gradient functional with an amount of predefined full-range exact exchange interaction of 0.25 and uses no cutoff in the calculation of Coulomb interaction.
 In the ground state it has an $e$ orbital partially filled by three electrons.
 The hole that represents the three electrons that occupy a fourfold degenerate spin-orbit state is dynamic Jahn-Teller active.
 We found the local minimum and saddle point geometrical configurations; however, in the optimization process the molecules experienced spurious translations and rotations. 
 To eliminate these numerical artifacts, we calculated the vibrational modes by means of the PBE functional and projected out the last six degrees modes associated with translations and rotations from atomic displacements.
 As a result, we found that the doubly degenerate orbital states of the $\text{CH}_3\text{O}$ interacts with an effective vibrational mode whose frequency is $130.50$~meV. 
 We report the Ham reduction factor as $p=0.47$.
 We note that a similar quenching factor ($d = 0.478$) was reported in Ref.~\cite{Sharma_2025}, where EOM-CCSDT/ANO1 level of theory \cite{eqmo} was applied to approximate the potential energy surface until the quartic order. 
 Moreover, we performed non-collinear calculation of the energy splitting due to spin-orbit coupling as described in Sec. \ref{sec:N3V}. However, we had to smear the three electrons that resides on the four spin-orbital states of the $e$ orbital by forced 0.75 occupations by $(e_{+\uparrow}^{0.75}e_{-\uparrow}^{0.75}e_{+\downarrow}^{0.75}e_{-\downarrow}^{0.75})$ occupation in order to avoid the disastrous effect of spin contamination.
 Otherwise, we experienced 3$\times$ underestimation upon calculating $\lambda_0$ within the $(e_{+\uparrow}^{1}e_{-\uparrow}^{1}e_{+\downarrow}^{0.5}e_{-\downarrow}^{0.5})$ constraint.
 Our intrinsic value of $\lambda_0 = 17.99$~meV is further reduced by the Ham factor that results in $\lambda_{\text{theory}}=8.47$~meV which is in good agreement with the experimental value in Ref.~\cite{CH3O_exp} see Table~\ref{tab:methoxy_split} where we additionally compare the higher level multimode case that we will discuss in the following section. 
 
\subsection{The general multimode treatment for $\mathrm{CH}_3\mathrm{O}^0$}
\label{sec:CH3O_multi}
We determine the $(A_{1}\oplus A_{1}\oplus...\oplus E\oplus E\oplus...)\otimes e$ multimode Jahn-Teller case where an arbitrary amount of vibration modes transforming as $E$ and $A_1$ irreducible representations that of $C_{3v}$ are being coupled to a Jahn-Teller unstable degenerate electronic multiplet spanned by $e$. 
We apply the Condon-Shortley convention that takes advantage of the orbital states $(\ket{e_x}, \ket{e_y})$ and phonon vibration $(\hat{q}_{ix}, \hat{q}_{iy})$ double degeneracy both on electronic ($|e_{\pm}\rangle=\mp(|e_{x}\rangle\pm i|e_{y}\rangle)/\sqrt{2}$) and nuclear $\hat{q}_{i\pm}=\mp(\hat{x}_{i}\pm i\hat{y}_{i})$ degrees of freedom. This convention allows us to express the polynomial basis of the phonon normal modes representation of nth order by simple monomials. Methoxy molecule has three modes transforming as $A_1$ $(\hat{q}_1,\hat{q}_2,\hat{q}_3)$ and three doubly degenerate modes $(\hat{q}_{4x/y},\hat{q}_{5x/y},\hat{q}_{6x/y})$. We designed our code so when the user describe phonon modes, using these convention, the code can parse polinomial expressions that contain the corresponding complex basis polynomials: $(\hat{q}_{i\text{+/-}})$. It is important to note that the user can easily include the spin-orbit coupling Hamiltonian by defining a zeroth order polynomial between the electron states.

\begin{figure}[h!]
\includegraphics[width=0.45\textwidth]{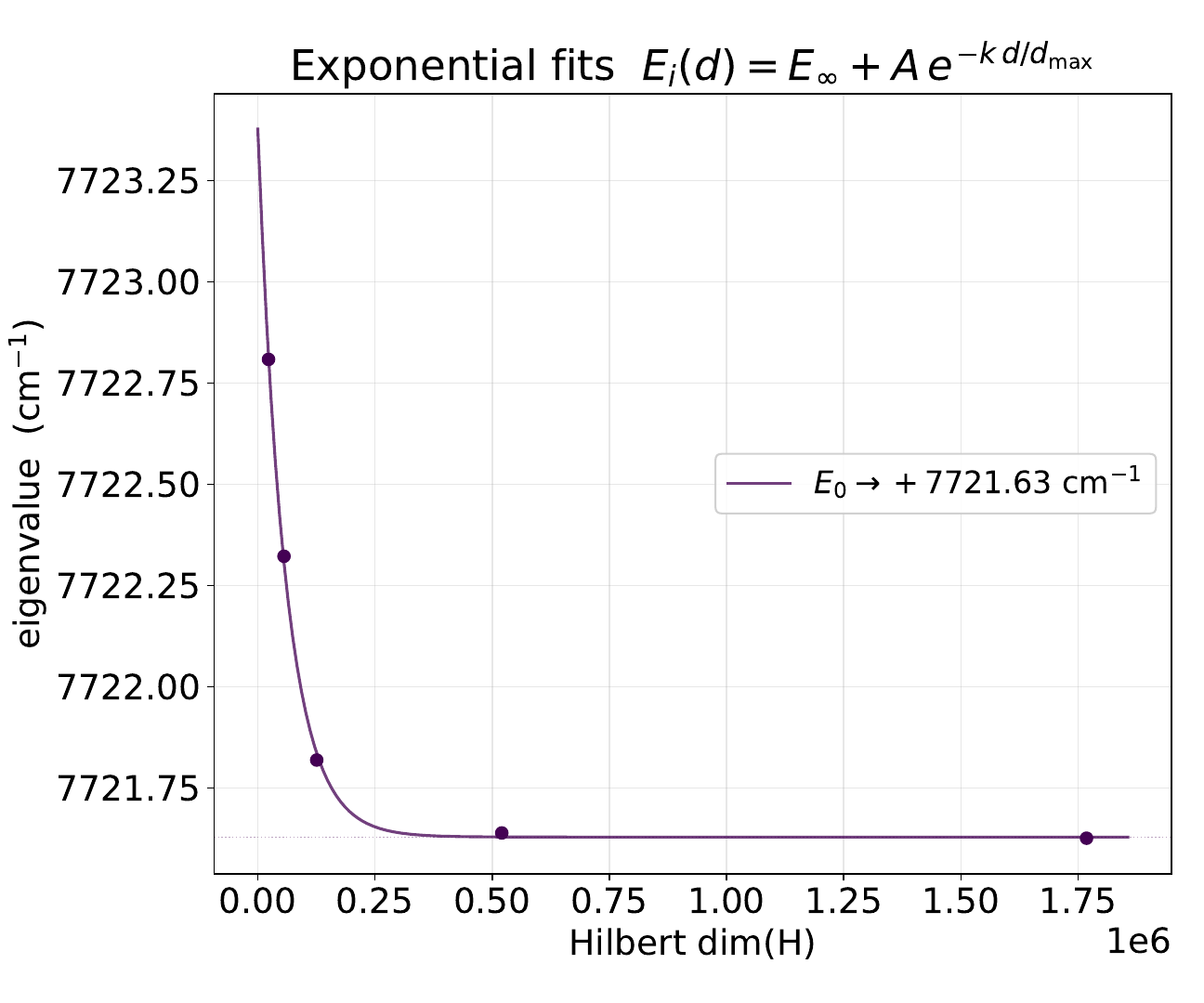}
\centering
\caption{This figure shows the calculated APES of the excited state of the neutral $ \text{N}_{3}\text{V}$ defect. The a) and b) part of the figure represents the $\text{N}_3\text{V}^0$ point defect.}
\label{fig:methoxy_gnd_conv}
\end{figure}

The fourth order dynamic Jahn-Teller Hamiltonian of the methoxy radical contains hundreds of terms. In order to simplify the inputs, and lower running time we set \texttt{jtd\_run} to symmetrize Hamiltonian and complete the diagonal terms. We calculated the 10 lowest vibrionic states of methoxy radical using various phonon energy cuts $(12000-24000\text{cm}^{-1})$ with Hilbert space dimensions 22882-1767930. We set the number of Lanczos iterations to 2000. We tested the convergence and found that by changing the phonon energy cut from $20000~\text{cm}^{-1}$ (520484 dimensions) to $24000~\text{cm}^{-1}$ (1767930 dimensions), the eigenvalues improve by less then $0.1~\text{cm}^{-1}$. Our vibrionic energy level values are presented in Table \ref{tab:methoxy_eigens} along with experimental values\cite{Sharma_2025,methoxy_exp_3,methoxy_exp_4,methoxy_exp_6,methoxy_exp_7,methoxy_exp_8}. Table shows the splitting of the first two energy levels using different methods. Figure \ref{fig:methoxy_gnd_conv} demonstrates how the methoxy's ground state is converging with respect to the Hilbert space dimension.

\begin{table}[!ht]
    \centering
    \caption{This table presents the eigenvalues of the 1767930 dimensional Hamiltonian that describe the methoxy radical, compared with the experimental data taken from \cite{Sharma_2025,methoxy_exp_3,methoxy_exp_4,methoxy_exp_6,methoxy_exp_7,methoxy_exp_8}.}
    \label{tab:methoxy_eigens}
    \begin{tabular}{lll}
    \hline
        eigenvalue  & energy level  & experiment \\
        $(\text{cm}^{-1})$ & $(\text{cm}^{-1})$ & $(\text{cm}^{-1})$\\
        \hline
        7721.63 & 0.00 & 0 \\ 
        7785.95 & 64.33 & 61.5 \\ 
        8417.04 & 695.41 & 689.3 \\ 
        8672.51 & 950.88 & 945.9 \\ 
        8771.57 & 1049.94 & 1041.8 \\ 
        8837.33 & 1115.71 & 1108 \\
        8961.15 & 1239.52 & 1224.6 \\ 
        8971.02 & 1249.39 & 1233.7 \\ 
        9080.51 & 1358.89 & 1346.8 \\ 
        9101.66 & 1380.04 & 1365.7 \\ 
        9144.70 & 1423.07 & 1411.3 \\ \hline
    \end{tabular}
\end{table}

\begin{table}[h]
    \centering
    \caption{Comparison of our single mode Jahn-Teller treatment on top of DFT calculations as described in Section \ref{sec:CH3O} to the multimode treatments that of Refs.~\cite{Sharma_2019, Sharma_2025, Weichman_2017} and the experimentally reported value $\lambda_{\text{exp}} = 7.94$~meV from Ref.~\cite{CH3O_exp}. 
    All values are in meV units while "p.w." refers to the present work. $\lambda_\mathrm{theory}$ values under () parentheses refers recalculated values by our $\mathtt{jahn {\text -} teller {\text -} dynamics}$ code.
    }
    \label{tab:methoxy_split}
    \begin{tabular}{lllll}
    \hline 
         & $^\mathrm{a}$ p.w.        & $^\mathrm{b}$      & $^\mathrm{c}$  & $^\mathrm{d}$  \\
        \hline
       $\lambda_0$     & 18.0 & 16.5 & 16.5 & 16.5 \\   
       $\lambda_\mathrm{theory}$  & 8.47 & 7.53 & 7.87 (7.80) & 7.69 \\ 
        \hline
    \end{tabular}
    \newline
    \caption*{{$^\mathrm{a}$ single mode $E\otimes e$ JT case, up to 2nd order terms }
    \newline
    {  $^\mathrm{b}$ multimode JT, up to 2nd order terms without cross-terms, see Supp. Mat. of Ref.~\cite{Sharma_2019} }
    \newline
    {  $^\mathrm{c}$ multimode JT, up to 4th order terms, see Supp. Mat. of Ref.~\cite{Sharma_2025} }
    \newline
    { $^\mathrm{d}$ multimode JT, up to 4th order terms, see Ref.~\cite{Weichman_2017} }}
\end{table}

The Taylor expansion of the coupling of electron orbitals and phonons is the same as in equation \eqref{eq:generic_e_ph3}. 
In order to express the electron-vibration interaction from Eq.~\eqref{eq:generic_e_ph3} in the circular basis for the first order:

\begin{equation}
\label{eq:methoxy1st}
    \begin{split}
        \hat{H}_{\text{1st}} = \underbrace{\ket{e_+}\bra{e_+}\sum_i F^{(++)}_{i}\hat{q}_i}_{\text{user input}} +
        \underbrace{\ket{e_-}\bra{e_-}\sum_i F^{(++)}_{i}\hat{q}_i^\dagger}_{\text{diagonal completion}} \\
        +\underbrace{ \ket{e_+}\bra{e_-}\sum_{i}F^{(+-)}_i\hat{q}_i}_{\text{user input}} + 
        \underbrace{\ket{e_-}\bra{e_+}\sum_{i}F^{(+-)*}_i\hat{q}_i^\dagger}_{\text{Hermitian completion}}
    \end{split}
\end{equation}
generally in the arbitrary order of methoxy's dynamical Jahn-Teller interaction $i$ runs over the real coordinates of three $A_1$ modes (1, 2, 3) and as well as three double degenerate $E$ modes (4+,4$-$,5+,5$-$,6+,6$-$), however in the first order, only 4$\pm$, 5$\pm$, 6$\pm$ coordinates are involved in the upper triangle of the Hamiltonian leading to a $(E\oplus E \oplus E )\otimes e$ three-mode JT case. Eq.~\eqref{eq:methoxy1st} shows how Hermitian completion diagonal completion works. The higher order terms are generated based of Eq.~\eqref{eq:gen_H}. The other option, what we used is to define the upper triangle of the system such that the offdiagonal coupling coefficients are multiplied by a factor of two, and then the  code symmetrize the Hamiltonian.

\begin{equation}
\label{eq:methoxy1stupper}
   \begin{split}\hat{H}_{\text{1st}}^{\text{upper}} & =\overset{\text{user input}}{\overbrace{|e_{+}\rangle\langle e_{+}|\sum_{i}F_{i}^{\text{++}}\hat{q}_{i}}}+\overset{\text{diagonal completion}}{\overbrace{|e_{-}\rangle\langle e_{-}|\sum_{i}F_{i}^{\text{--}}\hat{q}_{i}}}\\
   & +\underbrace{|e_{+}\rangle\langle e_{-}|\sum_{i}2\times F_{i}^{\text{+-}}\hat{q}_{i}}_{\text{user input}}\\
   \hat{H}_{\text{1st}} & =\dfrac{1}{2}\left(\hat{H}_{\text{1st}}+\hat{H}_{\text{1st}}^{\dagger}\right)
\end{split}
\end{equation}

Our $\mathtt{jahn {\text -} teller {\text -} dynamics}$ package can parse and interpret polynomials in any order in the circularly polarized basis ($\hat{q}_{i\pm}$, $|e_\pm\rangle$). The user can set Hermitian completion and diagonal completion which follows the symmetry arguments explained by the Supp. Mat. of Ref.~\cite{Sharma_2025}.
We demonstrate the capabilities of our code by loading up all polynomial expressions and dynamic Jahn-Teller coefficients up to quartic order directly obtained from the data presented in the Supp. Mat. that of Ref.~\cite{Sharma_2025} where six vibration modes (three non degenerate and three degenerate) are coupled to the electronic structure.


\section{Pseudo Jahn-Teller case of the butatrien molecule}

The pseudo Jahn-Teller effect occurs between electron orbital states that lies so close to each other that the nuclear vibrations are strong enough to create coupling between them. One remarkable example of this interaction is the butatrien cation $(C_4H_4^\text{+})$ that's absorption spectrum is a well studied phenomena. It's two lowest energy states $^2B_{3g}$ and $^2B_{3u}$ are coupled together\cite{butatrien70s,butatrien00s}. In this study we use vibrational modes and coupling coefficients of the two-mode linear model provided by \cite{butatrienFCs}'s Supplementary Material and set the maximum number of vibrational quanta $N = 40$. They do not consider the null point vibration in the vibrational energy Hamiltonian, and they use dimensional coordinate, therefore we implemented position operators in our codebase such that $(\hat{q}_i = \frac{1}{\sqrt{2\omega_i}}\left( \hat{a} + \hat{a}^{\dagger} \right))$, where $\omega_i$ is the energy quantum of the ith harmonic oscillator, that in this case represents nuclear motion in the butatrien cation. The pseudo Jahn-Teller interaction between $\ket{^2B_{3g}}$ and $\ket{^2B_{3u}}$ states can be written as

\begin{equation}
    \begin{split}
        H_{\text{pseudo}} &= \underbrace{\ket{^2B_{3g}}\bra{^2B_{3u}} \otimes F^{gu}_{1} \hat{q}_{1}}_{\text{user input}} +\underbrace{\ket{^2B_{3g}}\bra{^2B_{3g}} \otimes F^{gg}_{2} \hat{q}_{2}}_{\text{user input}}\\&+\underbrace{\ket{^2B_{3u}}\bra{^2B_{3g}} \otimes F^{gu}_{1} \hat{q}_{1}}_{\text{Hermitian completion}}  + \underbrace{\ket{^2B_{3u}}\bra{^2B_{3u}} \otimes F^{uu}_{2} \hat{q}_{2}}_{\text{user input}}
    \end{split}
\end{equation}
 unbraced terms are defined by the user.

\begin{figure}[h!]
\includegraphics[width=0.45\textwidth]{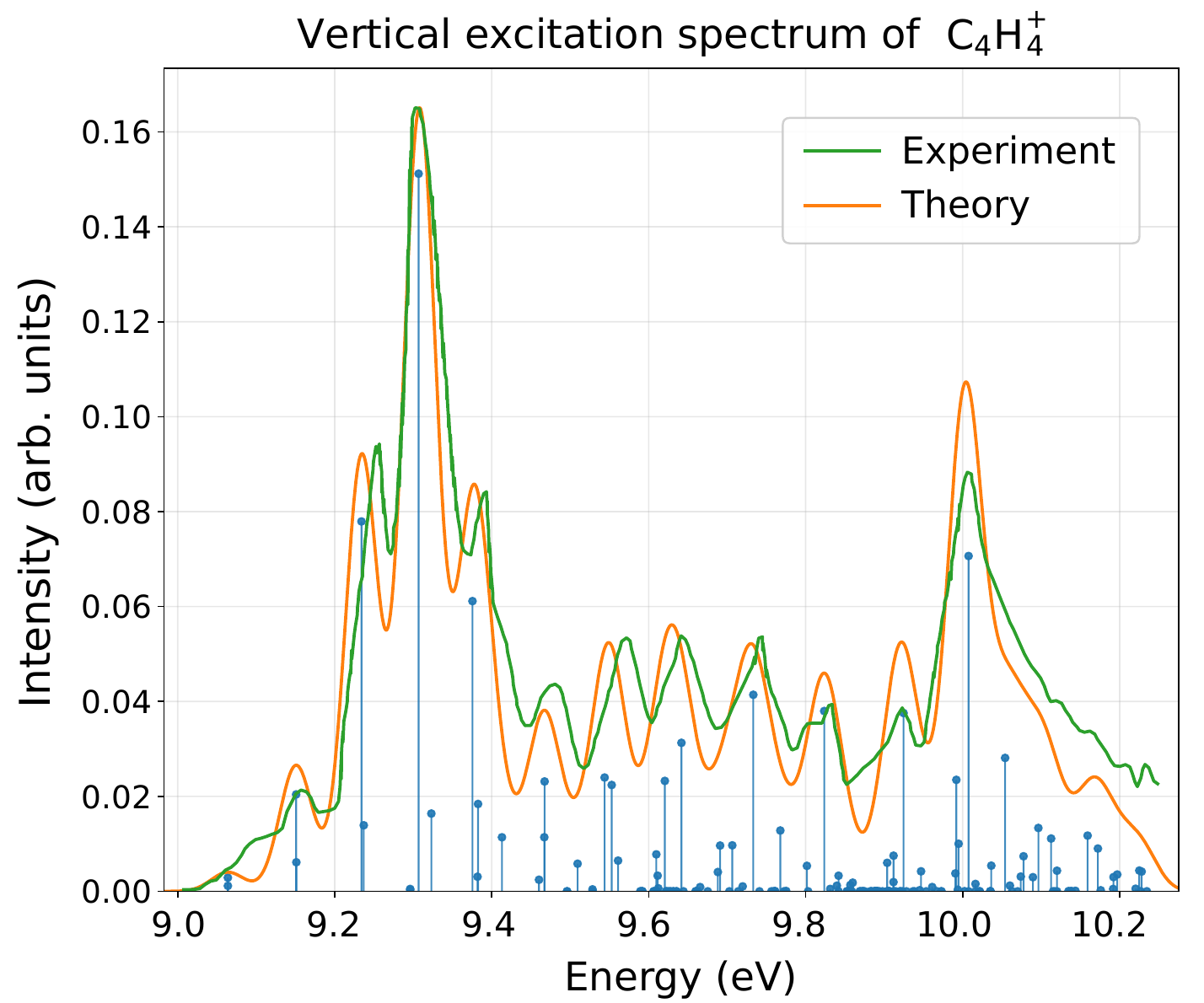}
\centering
\caption{The vertical spectrum of the butatrien cation $(C_4H_4^\text{+})$, Gaussians with $\text{fwhm} = 0.05$ is fitted to each points.}
\label{fig:butatrien}
\end{figure}

The vertical excitation spectrum can be seen in Figure \ref{fig:butatrien} experimental values are taken from \cite{butatrien_exp}.

\begin{equation}
        I(E_\eta) = |\braket{ \psi_0| \psi_i }|^2    \text{  where  } 
        \psi_0 = \dfrac{1}{\sqrt{2}}\left(\ket{^2B_{3g}, 0,0} + \ket{^2B_{3u}, 0,0}  \right)
\end{equation}
 and $\psi_\eta$ is the ith eigenstate of the system with energy $E_\eta$. \texttt{jtd\_run} provides the results in \texttt{eigenstates.npz}. It contains the basis states (\texttt{basis\_labels.npy}), the eigenvalues (\texttt{eigenvalues.npy}) and the eigenstates (\texttt{eigenvectors.npy}), therefore the user can easily postprocess the results to produce a vertical excitation or absorption spectrum.

\section{Summary}
\label{sec:summary}
In this work, we introduced $\mathtt{jahn-teller-dynamics}$, a Python-based library designed to model any dynamic Jahn-Teller systems. On top of this library we created $\mathtt{Exe.py}$, a Python script specifically designed to perform $E\otimes e$ Jahn-Teller active systems that experience spin-orbit coupling using outputs of density functional theory calculation. The code enables accurate calculation of energy splittings induced by vibronic and spin-orbit interactions and then simulates optical transition energies in the presence of an external magnetic field -- parameters of direct relevance to current experimental and theoretical studies in the recent years. Its capabilities were demonstrated by extracting key theoretical parameters for G4V centers in diamond, simulating the ZPL fine structure of the $\text{SnV}^-$ center, and we applied the code to predict the ZPL fine structure splitting energy of the neutral $\text{N}_3\text{V}^0$ defect in diamond and the reduced spin-orbit coupling in the methoxy ($\text{CH}_3\text{O}^0$) radical.

The general dynamic Janh-Teller modelling infrastructure is implemented in $\mathtt{jtd\_run}$. This script enables the user do define any type of dynamic Jahn-Teller system with various degenerate and non-degenerate electron states and vibrational modes. In order to present it's capabilites we calculated the energy levels of the methoxy $\mathrm{(CH_3O^0)}$ radical using the fourth order multimode dynamic Jahn-Teller model that was introduced by \cite{Sharma_2025}. Moreover, we calculated the vertical excitation spectrum of the butatrien cation $\mathrm{(C_4H_4^+)}$ that is a pseudo Jahn-Teller system. In all cases, the results show good agreement with experimental data. Due to it's user friendly design, it can be readily extended to treat more complex scenarios, such as the $T \otimes (t+e)$ Jahn-Teller system~\cite{Bhattacharyya_2013, Thiering_2023, PhysRevMaterials.6.034601, PhysRevB.95.014108}, pseudo Jahn-Teller effect~\cite{Bersuker_2013, PhysRevB.96.174105, PhysRevB.98.085207, Bersuker_2020}, and product Jahn-Teller systems~\cite{PhysRevB.65.035104, Thiering_2019, Ciccarino_2020, Bersuker_2020}. 

Therefore we can conclude that $\mathtt{jahn-teller-dynamics}$ is a robust codebase that can solve any dynamical Jahn-Teller problem and can serve as a useful asset for molecular and solid state physicists.



\section*{CRediT authorship contribution statement}
\textbf{Bal\'azs T\'oth:} Writing – original draft, Software, Validation, Methodology, Investigation, Conceptualization, Data curation;
\textbf{\'Ad\'am Gali:} Writing – review \& editing, Supervision, Funding acquisition
\textbf{Gerg\H{o} Thiering:} Writing – review \& editing, Supervision, Project administration, Formal
analysis, Conceptualization, Validation, Funding acquisition

\section*{Data availability}
The data that support the findings of this article will be openly available in the HUN-REN ARP database under the entry~\cite{ARP/EXJKGL_2025}. Additional exemplary data (NV center in diamond) can be found in the ARP database entry~\cite{ARP/O4CEI6_2025} of Ref.~\cite{thiering2024nuclear}.

\section*{Declaration of competing interest}
The authors declare that they have no known competing financial interests or personal relationships that could have appeared to influence the work reported in this paper.

\section*{Declaration of generative AI and AI-assisted technologies in the writing process}
The authors used the grammar assisting tools provided by Writefull (within Overleaf) and ChatGPT to improve the overall readability and the quality of the text and the styling of figures. 
The authors reviewed and edited the content from sentence to sentence and acted appropriately for every single suggestion made by these tools, and thus the authors take full responsibility for the content of the manuscript.

\section*{Acknowledgements}
The project supported by the Doctoral Excellence Fellowship Programme (DCEP) is funded by the National Research Development and Innovation Fund of the Ministry of Culture and Innovation and the Budapest University of Technology and Economics. 
Support by the Ministry of Culture and Innovation and the National Research, Development and Innovation Office (NKFIH) within the Quantum Information National Laboratory of Hungary (Grant No.\ 2022-2.1.1-NL-2022-00004) is much appreciated.
A.G.\ acknowledges the high-performance computational resources provided by KIF\"U (Governmental Agency for IT Development), the European Commission for the projects SPINUS (Grant No.\ 101135699) and QuSPARC (Grant No.\ 101186889) as well as the QuantERA II projects Maestro (NKFIH Grant No.\ 2019-2.1.7-ERA-NET-2022-00045) and Sensextreme (NKFIH Grant No.\ 2019-2.1.7-ERA-NET-2022-00040).
G.T.\ was supported by the J\'anos Bolyai Research Scholarship of the Hungarian Academy of Sciences and by NKFIH under Grant No.\ STARTING 150113.

\bibliographystyle{elsarticle-num}



\end{document}